# Sensitive Chronocoulometric Detection of miRNA at Screen-printed Electrodes modified by Gold decorated MoS$_2$ Nanosheets

*Abhijit Ganguly,[†] John Benson,[‡] and Pagona Papakonstantinou*[*,†]*

[†]School of Engineering, Engineering Research Institute, Ulster University, Newtownabbey BT37 0QB, United Kingdom

[‡]2-DTech, Core Technology Facility, 46 Grafton St., Manchester M13 9NT, United Kingdom

[*]Corresponding author, e mail address: *p.papakonstantinou@ulster.ac.uk*

**KEYWORDS.** *Molybdenum disulfide (MoS$_2$) Nanosheets, Ionic Liquid Assisted Grinding Exfoliation, Gold Decorated MoS$_2$ Hybrid Nanosheets, Screen-Printed Gold Electrodes, Electrochemical Deposition, microRNA-21 (miRNA-21), Electrochemical microRNA-sensor, Chronocoulometric Detection*

**ABSTRACT:** Developing novel simple and ultrasensitive strategies for detecting microRNAs (miRNAs) is highly desirable because of their association with early cancer diagnostic and prognostic processes. Here a new chronocoulometric sensor, based on semiconducting 2H MoS$_2$ nanosheets (MoS$_2$ NSs) decorated with a controlled density of monodispersed small gold nanoparticles (AuNPs@MoS$_2$), was fabricated via electrodeposition, for the highly sensitive detection of miRNA-21. The size and interparticle spacing of AuNPs was optimized by controlling nucleation and growth rates through tuning of deposition-potential and Au-precursor concentration and by getting simultaneous feedback from morphological and electrochemical activity studies. The sensing strategy, involved the selective immobilization of thiolated capture probe DNA (**CP**) at AuNPs and hybridization of **CP** to a part of miRNA target, whereas the remaining part of the target was complementary to a signaling non-labelled DNA sequence that served to amplify the target upon hybridization. Chronocoulometry provided precise quantification of nucleic acids at each step of the sensor assay by interrogating [Ru(NH$_3$)$_6$]$^{3+}$ electrostatically bound to phosphate backbones of oligonucleotides. A detailed and systematic optimization study demonstrated that the thinnest and smallest MoS$_2$ NSs improved the sensitivity of the AuNP@MoS$_2$ sensor achieving an impressive detection limit of ≈100 aM, which is 2 orders of magnitude lower than that of bare Au electrode and also enhanced the DNA-miRNA hybridization efficiency by 25%. Such improved performance can be attributed to the controlled packing density of **CP**s achieved by their self-assembly on AuNPs, large interparticle density, small size and the intimate coupling between AuNPs and MoS$_2$. Alongside the outstanding sensitivity, the sensor exhibited excellent selectivity down to femtomolar concentrations, for discriminating complementary miRNA-21 target in a complex system composed of different foreign targets including mismatched and non-complementary miRNA-

155. These advantages make our sensor a promising contender in the point of care miRNA sensor family for medical diagnostics.

- **INTRODUCTION**

Micro RNAs (miRNAs) is a class of short (about ~19–23 nucleotides) single-stranded non-coding RNAs that regulate gene expression and cellular processes.[1-3] Studies have demonstrated that abnormality in miRNA expression is closely related to initiation and progression of cancers. For example, overexpressed circulating level of miRNA-21 was considerably higher in plasma specimens of patients suffering from breast, cervical, lung or pancreatic cancer compared to healthy controls. As a result, miRNA-21 has become one of the clinically important diagnostic biomarkers for cancer screening and disease progression.[1] However, the relatively low level of miRNAs expression, their small size and their inherent degradable nature make direct quantification particularly challenging, necessitating the development of new platforms for their accurate and straightforward quantification in clinical samples. Among these, electrochemical-based platforms hold promise, due to their advantages of fast analysis, cost-effectiveness, and simplicity of operation.

It is well established that, the physical structure of a DNA probe layer immobilized on the electrode surface is critical in defining the overall performance of the sensor in terms of selectivity, sensitivity and reproducibility. Although the self-assembly of thiolated DNA at the surface of a gold electrode, exploiting the well-established Au-S chemistry, is a widely employed immobilisation approach,[4-5] it remains challenging to precisely control the orientation and conformation of surface-tethered oligonucleotides and finely tune the hybridization efficiency. Theoretical studies employing thiolated DNA on gold flat surfaces have predicted that efficient hybridization occurs with large inter-probe distances and upright conformations;[4-6] densely packed probe surfaces should be avoided as they restrict the accessibility of target DNA molecules due to steric effects.[5-6] In practice, the assembly of DNA is influenced by several factors including interactions between nitrogen atoms of DNA bases and the Au surface. Intuitively one would expect that, the surface coverage of DNA-probe recognition layer can be regulated through a controlled gold nanoparticle (AuNP) distribution of small particle size, narrow size variation and appropriate particle separation, instead of employing flat gold surfaces, with multiple anchoring points. Our work verifies this hypothesis, by controlling the size and interparticle spacing of AuNPs through judicious choice of electrodeposition conditions and by getting simultaneous feedback from morphological and electrochemical activity studies.

Molybdenum disulfide ($MoS_2$) is an important member of transition-metal dichalcogenides (TMDC), with unique layered structure, consisting of a single layer of Mo atoms sandwiched between two layers of S atoms in a trigonal prismatic arrangement. The weak Van der Waals interactions between the $MoS_2$ sheets make it possible to exfoliate the bulk $MoS_2$ to a few-layers or even to a single-layer crystalline sheet, via mechanical,[7-8] chemical routes or a combination of both.[9-14] The decoration of a few-layer $MoS_2$ nanosheets ($MoS_2$ NSs) with noble metal nanoparticles (NPs), such as Au, Ag, Pt, has become a popular and effective way to functionalize the 2D $MoS_2$-surface and enhance its sensing performance.[2-3, 14-18] So far, the application of $MoS_2$ or gold decorated $MoS_2$ NSs (AuNPs@$MoS_2$ NSs) for miRNA-21 detection is limited to a few studies, mainly classified to fluorescence-quenching,[12-13, 19-21] surface-enhanced Raman scattering,[17] and electrochemical[2-3, 15] based detection methods. Interestingly, in most of the previous studies on AuNPs@$MoS_2$ hybrids,[2-3, 11, 17-18, 22] the $MoS_2$ NSs were exfoliated via the popular lithium intercalation route.[9] This exfoliation approach results in $MoS_2$ layers of metallic phase, with a high population of defects in the basal plane,[11, 18, 23] which can act either as nucleation sites for the growth of high density of AuNPs with a large variance in particle size or as anchoring sites for non-specific adsorption. Uncontrolled AuNP growth on MoS$_2$ NSs is a limitation for its use in nucleic acid sensing, as it favors a highly packed assembly of DNA probe immobilization, which restricts the accessibility of target molecules. Sonication of bulk single crystals in appropriate solvents provides MoS$_2$ NS of semiconducting 2H phase.[12, 14, 16, 21, 24] However, their decoration with AuNPs via chemical reaction routes is limited at the edges, due to the absence of defects in the basal plane, restricting dramatically their use.[16, 25] Hence, alternative methods for the controlled synthesis of AuNPs on defect free MoS$_2$ NSs should be sought; however this area is an almost unexplored terrain. In this contribution, we show that these requirements can be met under well controlled electrochemical deposition (ECD) conditions.[15, 26-31] ECD is also free of critical drawbacks such as the formation of "free" AuNPs, which usually coexist with AuNPs@MoS$_2$ hybrids in solution-based routes.[11, 22, 24, 32-33]

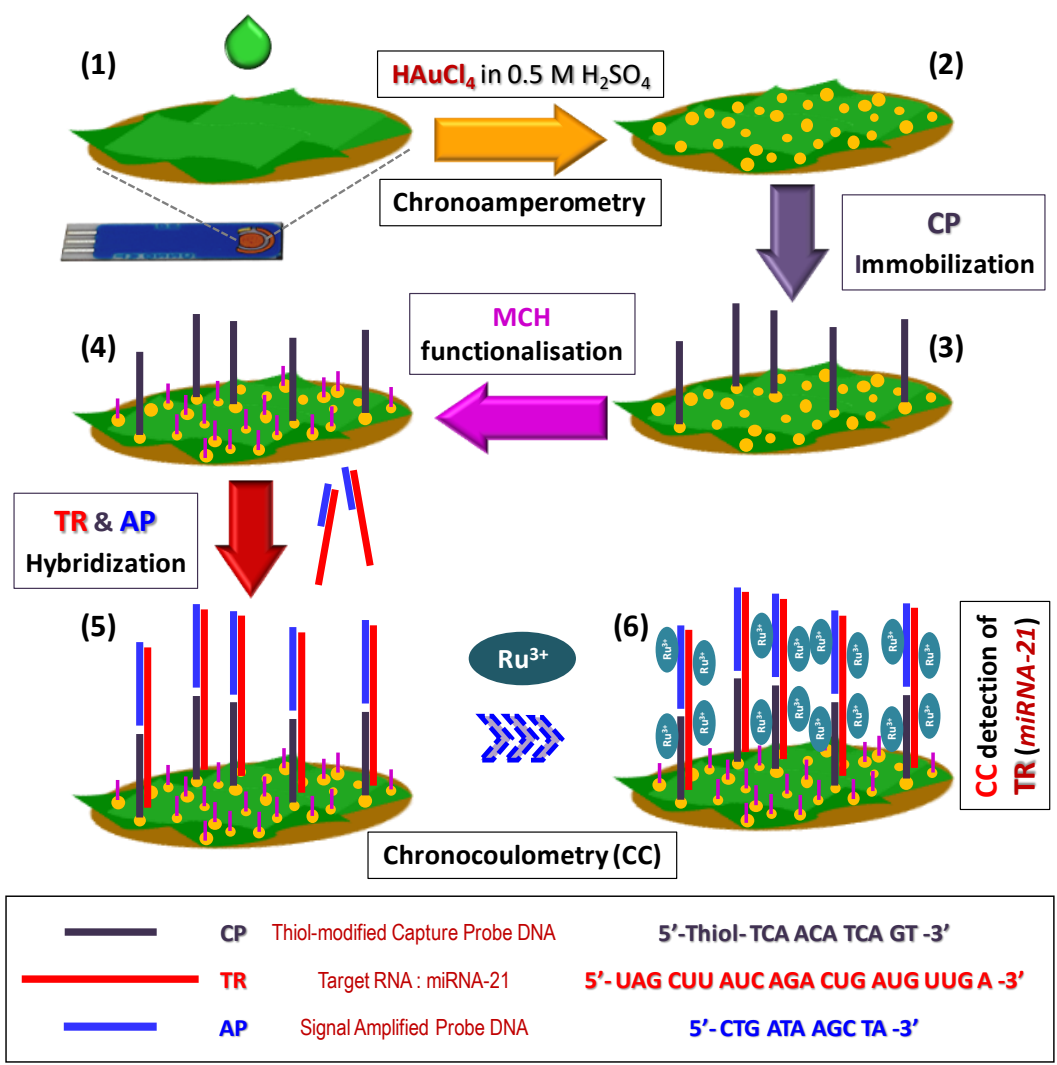

**Figure 1. Fabrication of miRNA-21 sensor and detection strategy.** Coating of MoS$_2$ nanosheets (MoS$_2$ NSs) on commercial screen-printed gold electrodes (SPGEs) (1); decoration of MoS$_2$ NSs with AuNPs (to create AuNPs@MoS$_2$ NSs) following an optimized chronoamperometric (CA) route (2); assembly of ssDNA capture probe: anti-miRNA-21 (**CP**), miRNA-21 target (**TR**) and signal amplification probe (**AP**) on AuNPs@MoS$_2$/SPGE sensor (3-5); chronocoulometric (CC) detection of miRNA (**TR**) by monitoring [Ru(NH$_3$)$_6$]$^{3+}$ (RuHex) electrostatically bound to phosphate backbones of oligonucleotides (6).

Herein, we report a simple and sensitive electrochemical platform for miRNA-21 detection using a screen-printed gold electrode (SPGE) modified with MoS$_2$ NSs decorated with a controlled density of monodispersed AuNPs (AuNPs@MoS$_2$ NSs) achieved by chronoamperometric (CA) electrodeposition. Utilizing SPGPE as a base platform offers the possibility of building a portable and disposable miniaturized electrode system suitable for both electrodeposition of AuNPs and subsequent bio-functionalization, for laboratory and onsite-clinical-measurements. Figure 1 illustrates schematically the fabrication of the sensor and the detection strategy (described in the *Supporting Information* (SI), Section S2), which involves selective immobilisation of thiolated capture probe ssDNA (**CP**) at AuNPs@MoS$_2$ and hybridisation of the immobilized **CP** to a part of miRNA target (**TR**), whereas the remaining part of **TR** is complementary to a ssDNA sequence (**AP**; Amplification Probe) that serves to amplify the hybridization signal (Table S1). We employ chronocoulometry to quantify the amount of nucleic acids at each step of the detection strategy by monitoring [Ru(NH$_3$)$_6$]$^{3+}$ (RuHex) electrostatically bound to phosphate backbones of DNA or DNA-RNA hybrids. A detailed optimization study on both AuNP deposition and immobilization steps achieved an impressive detection limit of ~100 aM, which is 2 orders of magnitude lower than that of bare Au electrode and also enhanced the DNA-miRNA hybridization efficiency by 25%. Moreover, this newly developed biosensor was highly specific toward the target sequence miRNA-21 demonstrating the ability to differentiate between sequences that differed even by a single base, along with a clear distinction in a medium consisting of many interfering targets mixed together.

Among the electrochemical based detection routes, we have employed the chronocoulometry (CC) technique as the detection method of choice, first proposed by Steel et al.[34] Literature reports have revealed that CC can be highly effective compared to any voltammetric methods in discriminating against background contribution at relatively high potentials; hence it can be used to generate a significantly more intense signal with higher resolution.[4, 34-35] CC is a fast (hundreds of milli-seconds) and non-destructive technique, particularly useful for analytes like DNA or RNA, which are prone to degradation even in relatively mild environments. So far, the principle and employment of CC based sensing approach has been established and optimized for DNA on planar Au electrode systems.[4, 6, 34-35] The work presented here is the first study that provides a detailed account on the optimization and employment of CC technique for the detection of miRNA on semiconductor/AuNP sensor.

- **RESULTS AND DISCUSSION**

**Electrochemical Deposition of Gold Nanoparticles on MoS$_2$ Nanosheets.**

Our initial studies revealed that the electrochemical pretreatment of the MoS$_2$/SPGE working electrode in H$_2$SO$_4$ was crucial for achieving well-controlled reproducible AuNP deposition. (SI, Section S3, Figures S2). Similarly, it was evident from our studies that the electrodeposition process is better controlled under a static applied potential (CA route) than that under dynamic potential scan (like CV method), resulting in homogeneous distribution of small AuNPs (SI, Section S4, Figure S3). We investigated the effects of applied potential and concentration of HAuCl$_4$ solution on the CA process as described below.

**Effect of Applied Potential ($V_{app}$).** The applied potential, $V_{app}$, is considered one of the main factors that govern the Au electrodeposition process, affecting the size and monodispersity of the resultant nanoparticles. It has been established, that negative overpotentials relative to standard potential for $AuCl_4^-$ reduction to $Au^0$, (+0.8 V versus Ag/AgCl (sat. KCl)) favor the creation of new nucleation sites over the growth of previously created nuclei.[26-29] Figures 2(a-d) demonstrate the effect of $V_{app}$ on the AuNPs electrodeposited on $MoS_2$/SPGE, when varying the $V_{app}$ potential from +0.1 to -0.2 V. It is observed that the application of decreasing $V_{app}$ potentials, leads gradually to a larger number of nucleation sites and hence to larger density of smaller, evenly distributed AuNPs, corroborating earlier work.[27-28]

Two growth modes are evident, depending on the $V_{app}$ potential. For positive $V_{app}$ values (+0.1V, 0V), the growth of a small number of initial nuclei is favored over the establishment of new nucleation sites as indicated by the large particle size and low packing density in Figures 2(a-b), consistent with earlier reports.[27-28] Statistical analysis, performed on the SEM images, presented by the particle size histogram in Figure 3a, confirms the observation. AuNPs grown at +0.1 V (least negative overpotential) result in large agglomerated particles (mean diameter, $D_m \approx$ 262 nm) with low particle density ($N_D$: particle number density $\approx 1.18$ particles/$\mu m^2$). Notably, the large standard deviation of NP-size, ($SD > \pm 108$ nm) represents an irregular size distribution of AuNPs, and the presence of aggregated Au particles.

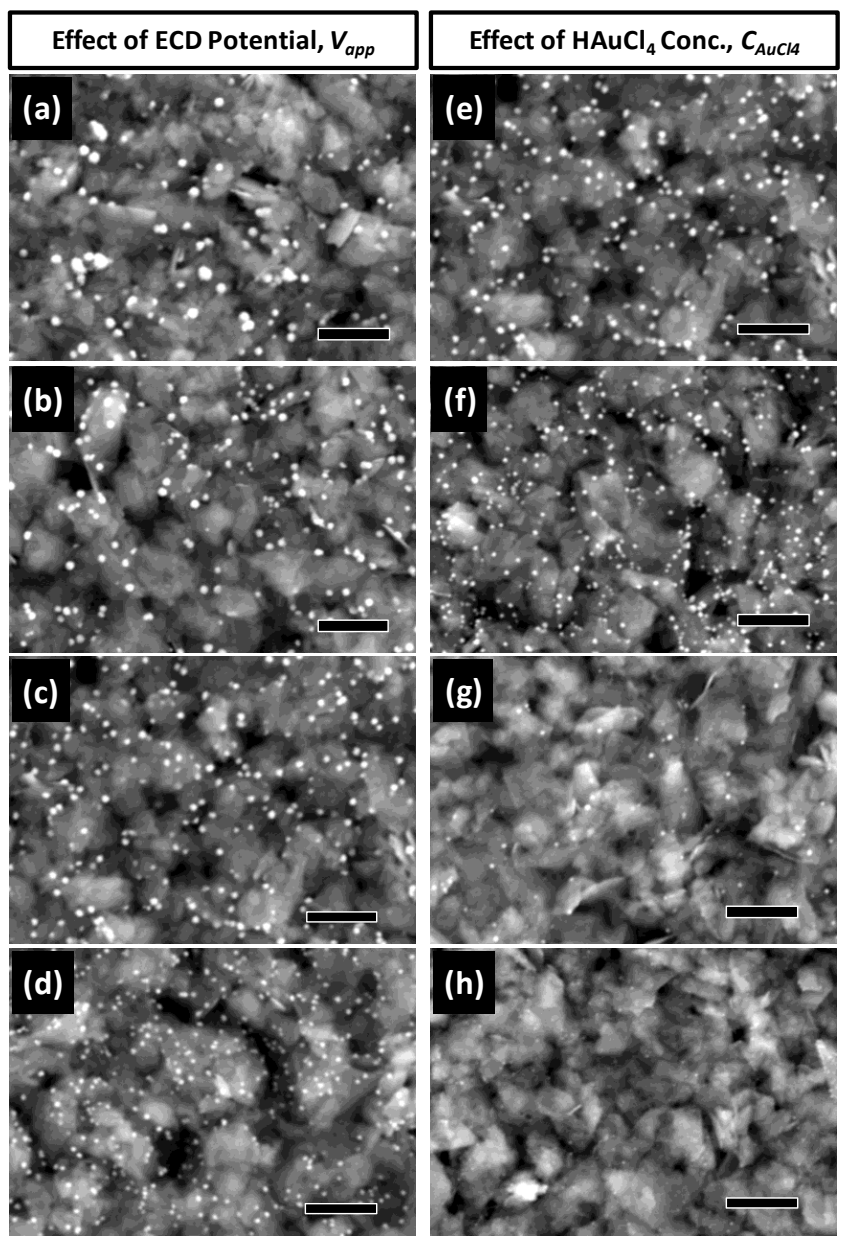

**Figure 2. Morphological characterisations of AuNPs@MoS₂ hybrids.** SEM images of AuNPs electrodeposited via chronoamperometry (CA) on the MoS$_2$/SPGEs electrodes (a-d) at different $V_{app}$: [(a) +0.1 V, (b) 0.0 V, (c) -0.1 V, (d) -0.2 V] with $C_{AuCl_4^-}$: 1 mM; and (e-h) at different $C_{AuCl_4^-}$: [(e) 1.0, (f) 0.5, (g) 0.1, and (h) 0.05 mM] at $V_{app}$: -0.1 V. $V_{app}$: Applied Potential,

$C_{AuCl_4^-}$: Concentration of HAuCl4 in 0.5 M H2SO4. MoS2 Loading: 50 µg. Potential scan-duration: 360 s. Scale bar: 1µm.

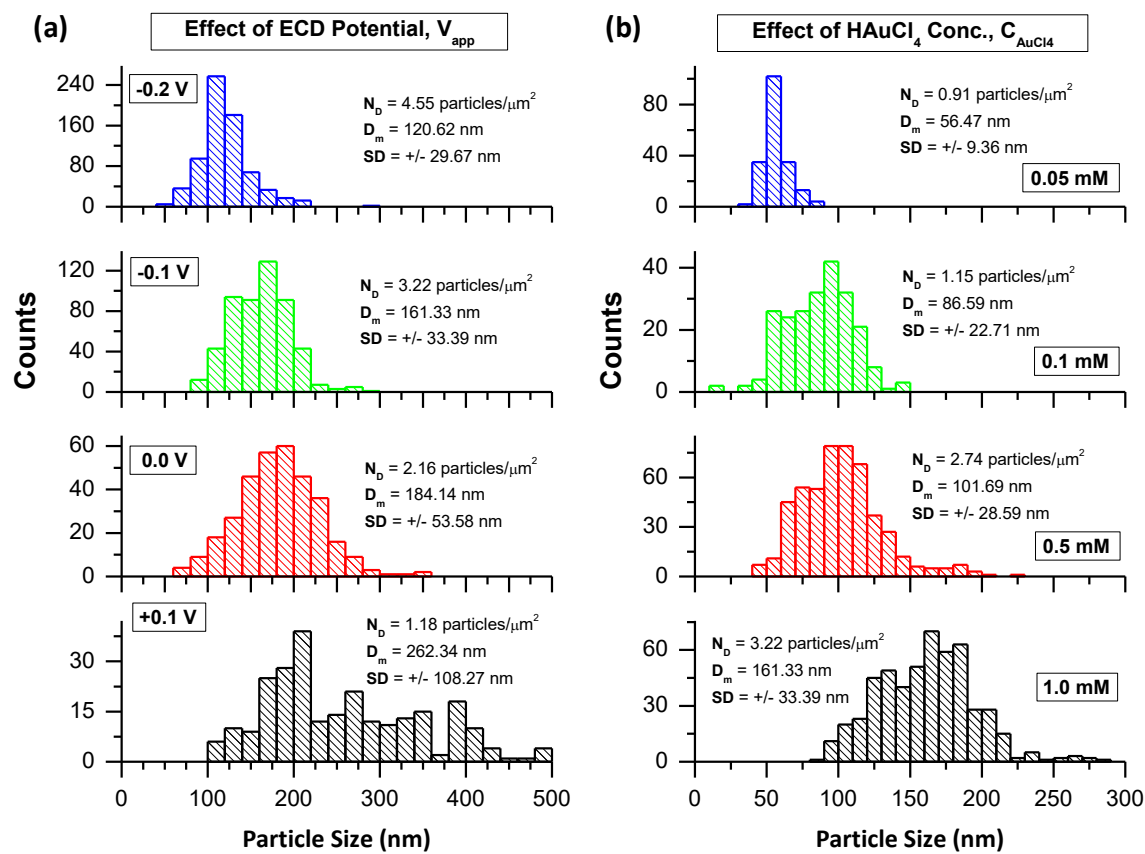

**Figure 3. Statistical analysis of morphological parameters.** Particle size histograms of AuNPs electrodeposited via CA on the MoS$_2$/SPGEs electrodes at various (a) Applied Potentials, $V_{app}$ (with $C_{AuCl_4^-}$: 1 mM), and (b) HAuCl$_4$ Concentrations, $C_{AuCl_4^-}$: (at $V_{app}$: -0.1 V). Statistical analysis has been performed on at least 3~4 independent SEM images for each sample. The statistical parameters of each sample are mentioned in its respective histogram: $N_D$: Particle Number Density, $D_m$: Mean Diameter of AuNP, and $SD$: Standard Deviation of NP-size.

In contrast, the application of negative $V_{app}$ values (-0.1V, -0.2V) favors instantaneous nucleation, facilitating the formation of higher density small size particles (Figures 2c-d). The particle size histogram (Figure 3a) reveals that the Au-ECD at -0.2 V could reduce the size ($D_m \approx$ 120 nm) of AuNPs by more than 2 times and improve the $N_D$ (~4.55 particles/μm$^2$) by nearly 4 times. Notably, the **SD** value drops down to ±30 nm indicating reduced size dispersion.

On the other hand, at augmented negative potentials, such as $V_{app}$ = -0.3 V, the AuCl$_4^-$ → Au$^0$ reduction rate became high enough, resulting in aggregation (SI, Section S5, Figure S5).

**Effect of HAuCl₄ Concentration ($C_{AuCl_4^-}$).** As discussed previously in order to produce a high particle density, a large negative nucleation overpotential should be applied. This results in a high nucleation density but also a fast growth rate.[30-31] Fast growth rate is problematic as it results in a rapid expansion of the diffusion zone around the growing nucleus. Diffusion zone is the area of electrolyte around the nucleus that has a reduced concentration of metal ions present compared to the bulk electrolyte, because ions are been reduced and merged into the growing nuclei. As these diffusion zones expand, adjacent zones eventually merge. Nuclei, whose diffusion zones have coupled, experience retarded growth compared to nuclei with isolated diffusion zones. As a result, diffusion zone coupling results in different growth rates, hence a range of particle sizes is being created. A large particle size distribution can result in considerable problems in terms of sensitivity and reproducibility of sensor performance. So efforts to eliminate the depletion region around each nucleus would eliminate interparticle diffusion coupling and hence allow the formation of nanoparticles with uniform size. Our strategy to keep diffusion zone coupling to a minimum involves slow growth rate via low concentrations of the gold precursor.

The radius of the depletion region around each nucleus varies proportionally with the bulk concentration of metal ions (here, AuCl$_4^-$).[26, 30-31] High concentrations of HAuCl$_4$ ($C_{AuCl_4^-}$), give rise to fast growth rate, thereby permitting inter-particle diffusion zone coupling to occur, resulting in various size distributions. In contrast, at lower concentrations of HAuCl$_4$, growth rate of nuclei is slowed, thereby keeping diffusion zone coupling to a minimum, thus leading to the formation of smaller diameter monodisperse AuNPs.

Figures 2(e-h) demonstrate the effect of $C_{AuCl_4^-}$ on the AuNPs electrodeposited on MoS$_2$/SPGE at an applied voltage of -0.1V. As $C_{AuCl_4^-}$ decreases, SEM images display progressively smaller AuNPs with lower packing density and improved size distribution, with the best values attained at $C_{AuCl_4^-} \approx 0.1 - 0.05$ mM (Figures 2g and 2h). The particle size histogram plots (Figure 3b) reveal that by reducing the $C_{AuCl_4^-}$ from 1.0 to 0.05 mM, the AuNPs size (*D$_m$*) reduces by nearly 3 times, while the *SD* value drops down to only ±9.36 nm, reflecting an improved degree of monodispersion on the resultant AuNPs.

**Electrochemical Characterization of AuNPs@MoS$_2$ Hybrid Nanosheets.** The effects of $V_{app}$ and $C_{AuCl_4^-}$ were further characterized (Figure 4) in order to optimize the AuNP electrodeposition strategy, by estimating two electrochemical parameters *Q$_{dp}$* and *Q$_{ox}$*. Here, *Q$_{dp}$* represents the total charge involved during the Au(III) reduction to AuNPs formation,[28-29] estimated by integrating the corresponding current transient curves obtained by CA (SI, Section S6, Figure S6a). The electrodeposited AuNPs are further characterized by performing CV experiments in a 0.5 M $H_2SO_4$ solution (Figure S6b),[29] The **$Q_{ox}$** denotes the charges related to the reduction of Au-oxides, estimated by integrating the area under the corresponding reduction peak.[28-29]

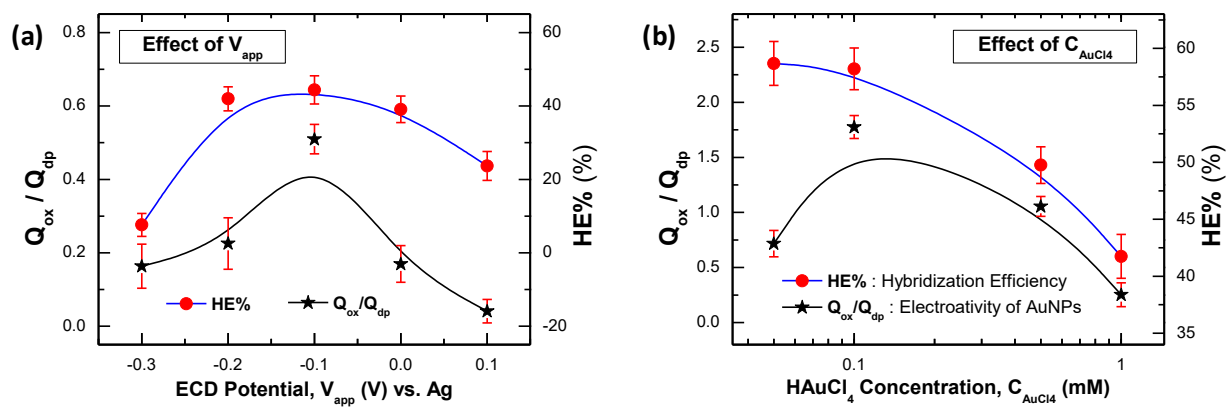

**Figure 4. Electrochemical characterisations of AuNPs@MoS$_2$ electrodes.** Effects of (a) applied potential ($V_{app}$) and (b) HAuCl$_4$ concentration ($C_{AuCl_4^-}$) on $Q_{ox}/Q_{dp}$ ratio and *HE%*. The $Q_{ox}/Q_{dp}$ ratio represents the yield of electro-active AuNPs for the AuNPs@MoS$_2$ hybrids. The *HE%* = [($\Delta Q$ × 100) / $Q_{CP}$] represents the hybridization efficiency, where $\Delta Q$ = [$Q_{CP-TR-AP}$ − $Q_{CP}$]. Error bars represent the standard deviations estimated from at least three independent measurements.

Consequently, the ratio of $Q_{ox}/Q_{dp}$ would represent the yield of electrochemically active AuNPs electrodeposited on the MoS$_2$/SPGE. As noted by Hezard et al.,[29] the Faradaic yield for Au electrodeposition should follow the relation: $\eta_{AuOx}/\eta_{Au} = 1.5 \times Q_{ox}/Q_{dp}$; the factor of 1.5 arises because 3 electrons are exchanged during Au(III) reduction, while only 2 electrons are involved in Au-oxide formation. Figures 4a and 4b demonstrate the dependence of $Q_{ox}/Q_{dp}$ ratio on $V_{app}$ and $C_{AuCl_4^-}$ respectively.

As shown in Figure 4a, AuNPs@MoS$_2$ hybrids electrodeposited at a high cathodic overpotential, with $V_{app}$ = -0.1 V, exhibiting uniformly distributed, relatively monodisperse small AuNPs ($D_m \approx$ 161 nm with $SD$ of ±33 nm; Figure 3a), possess the highest value for the $Q_{ox}/Q_{dp}$ ratio. Nevertheless, it is quite surprising that the AuNPs electrodeposited at $V_{app}$ = -0.2 V exhibit a fall on $Q_{ox}/Q_{dp}$ ratio, even though they possess a higher density and smaller particle size ($N_D \approx$ 4.6 particles/μm$^2$, $D_m \approx$ 120 ± 30 nm; Figure 3a) than those deposited at $V_{app}$ = -0.1 V ($N_D \approx$ 3.2 particles/μm$^2$). The lower $Q_{ox}$, hence the lower $Q_{ox}/Q_{dp}$, suggests instability of the deposited nuclei. It can be deduced that the as-deposited nuclei could be rearranged, while oxidized or even dissolved back in the solution before their reduction takes place during the backward scan.

Figure 4b illustrates that the effect of $C_{AuCl_4^-}$ on $Q_{ox}/Q_{dp}$ ratio. As $C_{AuCl_4^-}$ decreases, the $Q_{ox}/Q_{dp}$ becomes highest at $C_{AuCl_4^-}$ = 0.1 mM. Further reduction of $C_{AuCl_4^-}$ (= 0.05 mM) lowers the $Q_{ox}/Q_{dp}$ since it suffers seriously from very low ECD-yield of AuNPs. Based on the above results the optimized Au ECD conditions of the AuNPs@MoS$_2$ based sensor are as follows: $V_{app}$ = -0.1 V, $C_{AuCl_4^-}$ = 0.1 mM and MoS$_2$ loading = 50 μg.

**Elemental Characterisation of AuNPs@MoS$_2$ Hybrids.** The elemental characterization of AuNPs@MoS$_2$ hybrids was performed via X-ray photoelectron spectroscopy (XPS), and high-resolution spectra are illustrated in Figure 5 and Figure S8 (SI, Section S7). The wide survey scan of AuNPs@MoS$_2$ hybrid exhibits characteristic peaks for the main elements of Mo, S and Au. In addition to those elements, the presence of C and O elements is evident, originating from the solvent and the atmosphere. Calculated from the integrated areas of respective high resolution XPS spectra, the stoichiometric ratio of Mo to S was found to be close to 1:2 (1 : 2.10 ± 0.038), demonstrating the expected MoS$_2$ phase.

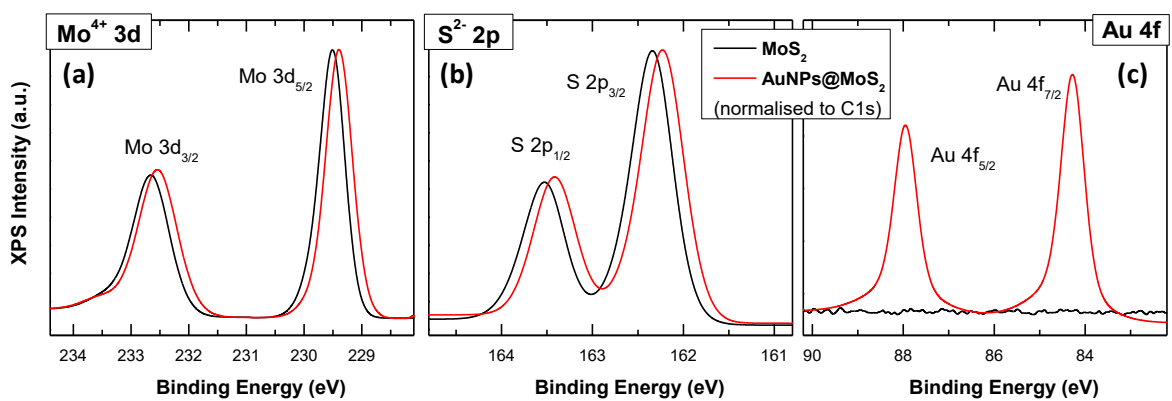

**Figure 5. Elemetal characterisation of AuNPs@MoS$_2$ hybrids nanosheets.** High-resolution XPS spectra of AuNPs@MoS$_2$ and pristine MoS$_2$ NSs drop-casted on SPGEs: (a) Mo$^{4+}$ 3d, (b) S$^{2-}$ 2p, and (c) Au 4f. All spectra are corrected by Shirley background and calibrated with reference to the C 1s line at 284.5 ± 0.2 eV associated with graphitic carbon. For AuNPs@MoS$_2$ hybrid NSs, the AuNPs is electrodeposited via CA for 360 s: $V_{app}$: -1.0 V. $C_{AuCl_4^-}$: 0.1 mM HAuCl$_4$ in 0.5 M H$_2$SO$_4$. MoS$_2$ Loading: 50 μg.

For AuNPs@MoS$_2$, the Mo 3d XPS spectrum (Figure 5a) shows doublet peaks at 229.38 and 232.53 eV attributed to Mo$^{4+}$ 3d$_{5/2}$ and Mo$^{4+}$ 3d$_{3/2}$ orbitals, respectively. Similarly doublet peaks around 162.34 and 163.53 eV, observed in Figure 5b, belong to S$^{2-}$ 2p$_{3/2}$ and S$^{2-}$ 2p$_{1/2}$ orbitals, respectively. These peak positions are indicative of Mo$^{4+}$ and S$^{2-}$ oxidation states in 2H phase of pristine MoS$_2$ NSs,[7, 36] indicating that the hybridization of MoS$_2$ NSs with AuNPs does not affect the crystallinity and chemical stability of MoS$_2$. Figure 5c shows the Au 4f spectrum, with doublet peaks positioned around 84.28 eV (Au 4f$_{7/2}$) and 87.95 (Au 4f$_{5/2}$), providing direct evidence for the reduction of the Au-precursors and hence the formation of AuNPs on MoS$_2$ NSs.[8, 36]

Interestingly, in the AuNPs@MoS$_2$, both Mo$^{4+}$ and S$^{2-}$ peaks (Figures 5a and 5b) exhibit an obvious red-shift to lower binding energies compared to that of pure pristine exfoliated MoS$_2$ NSs, indicating a down-shift of the Fermi level in MoS$_2$ due to p-type doping.[37-38] Here, the AuNPs act as a p-type dopant in MoS$_2$ since the AuCl$_4^-$ ions in solution can strongly withdraw electrons from MoS$_2$ layers and reduce to AuNPs.[23, 37-38]

**Electrochemical Optimization Studies for miRNA-21 detection.**

Initially the hybridization of the capture probe (**CP**) with the target miRNA sequence (**TR**), on the fabricated the AuNPs@MoS$_2$/SPGE platform, was confirmed by the presence of well resolved voltammograms of methylene blue redox signal, which was used an electrochemical indicator (SI, Section S8).

**Optimization of RuHex Concentration ($C_{RuHex}$): Adsorption Isotherm of RuHex.** A necessary and crucial step of the chronocoulometric detection was the determination of RuHex concentration at which the saturation condition could be achieved. At saturation condition, a complete charge compensation of the phosphate residues by redox cations was achieved i.e. one $[Ru(NH_3)_6]^{3+}$ cationic redox marker was electrostatically trapped for every three nucleotide phosphate groups.[34]

The influence of RuHex concentration ($C_{RuHex}$) at the **CP**-MCH-electrodes is shown in Figure 6. It is observed that the charge of surface-adsorbed RuHex, $Q_{ad}$ in the presence of **CP**, initially increases with $C_{RuHex}$ reaching saturation at $C_{RuHex} \geq 10$ μM on AuNPs@MoS$_2$/SPGEs (Figure 6a). Interestingly, at bare SPGEs, the adsorption saturation of RuHex is achieved at $C_{RuHex} \geq 40$ μM, which agrees reasonably well with the reported literature.[4, 6, 34-35, 39] The lower saturated charge values $Q_{ad}$ at AuNPs@MoS$_2$/SPGEs, compared to SPGE, can be understood in terms of the lower and controlled attachment of the thiolated **CP** on AuNPs leading to a lower negative charge density.

Adsorption isotherms for RuHex, at both AuNPs@MoS$_2$/SPGEs and bare SPGEs in the presence of **CP**, are presented in Figure 6b, satisfying the Langmuir adsorption model[34, 39] (SI, Section S9). From the linear fitting, the saturated coverage $Q_{sat}$ values are estimated as 1.73 and 3.68 μC, for the AuNPs@MoS$_2$/SPGEs and SPGE. Correspondingly, the estimated values of surface coverage density of **CP** probes ($\Gamma_{CP} = \Gamma_{DNA} = \Gamma_0(z/m)N_A$) for the **CP**-MCH-functionalized SPGE sensor agrees reasonably well with the reported values ($\Gamma_{CP} \approx 1 - 10 \times 10^{12}$ molecules/cm$^2$) for the shorter DNA-SAMs at the Au-electrodes.[4, 6, 34-35, 39] The 2-times higher $Q_{sat}$ value at the bare SPGE ($\Gamma_{CP} \approx 3 \times 10^{12}$ molecules/cm$^2$) indicates almost 2-times higher $\Gamma_{CP}$ values, compared to the AuNPs@MoS$_2$/SPGEs ($\Gamma_{CP} \approx 1.4 \times 10^{12}$ molecules/cm$^2$).

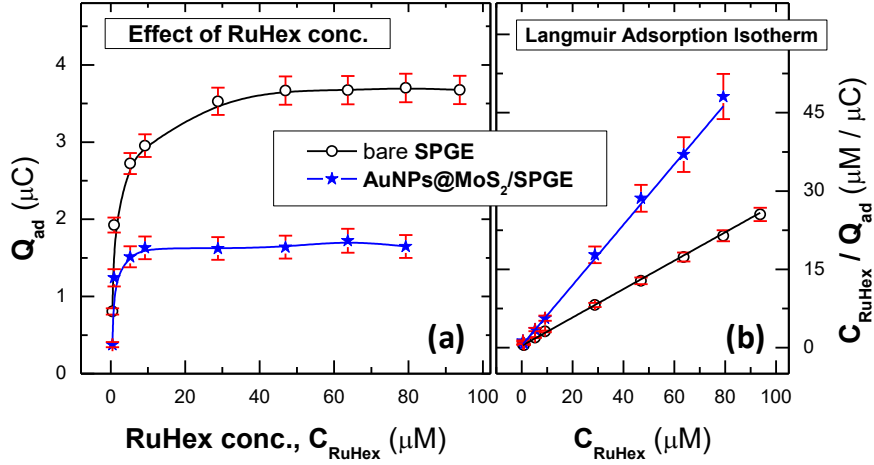

**Figure 6. Effect of RuHex concentration ($C_{RuHex}$)** and **optimization of CC-detection.** (a) Adsorption isotherm of RuHex for capture DNA probe (**CP**) immobilized on bare and AuNPs@MoS$_2$ coated SPGEs. $Q_{ad} = Q_{CP} = (Q_{RuHex} - Q_{blank})$ and $C_{RuHex}$ = RuHex concentration. (b) Plots of $C_{RuHex}/Q_{ad}$ versus $C_{RuHex}$, demonstrating the linear fitting of the binding data to the Langmuir adsorption isotherm.

Furthermore, the linear fitting of the isotherms (Figure 6b) reveals that the association constant **K** of RuHex at the AuNPs@MoS$_2$/SPGEs (1.43 µM$^{-1}$) is nearly double in magnitude than that (0.83 µM$^{-1}$) at the bare SPGEs. The observation suggests that association of [Ru(NH$_3$)$_6$]$^{3+}$ with **CP** improves significantly at the hybrid electrode. This finding is directly related to $\Gamma_{CP}$. At bare SPGE the high $\Gamma_{CP}$ regimes suffer from steric hindrance, which translates into a weaker binding affinity of cationic RuHex redox complex for DNA probe. This explanation is consistent with the binding constants values reported in literature, which are either slightly weaker for dsDNA ($K$ = 1.3 µM$^{-1}$) than for ssDNA ($K$ = 2.2 µM$^{-1}$) or are weaker for longer length DNA compared to shorter length DNA.[34, 39-40]

Based on our results presented in Figure 6, it can be concluded that a $C_{RuHex}$ ≥ 10 µM (14.5 µM was chosen) is sufficient for the CC detection of miRNA at the AuNPs@MoS$_2$/SPGEs, whereas for bare SPGEs-based sensors a $C_{RuHex}$ ≈ 50 µM is appropriate.

**Optimization of Sensing Strategies.** The miRNA detection as outlined in schematic diagram in Figure 1 involves the following optimization protocols: i) electrodeposition conditions of AuNPs, ii) concentration of capture DNA probes ($C_{CP}$), and iii) hybridization strategies of **CP** with the target miRNA (**TR**) and the signal amplified DNA probe (**AP**). Overall, the CC detection performance is defined by the hybridization efficiency *HE%* values, as the signature of the **CP**–**TR**–**AP** hybridization effectiveness.

**Effect of Au ECD parameters.** Figures 4a and 4b display the dependence of *HE%* value on applied potential ($V_{app}$) and HAuCl$_4$ concentration ($C_{AuCl_4^-}$), respectively. The effect of $V_{app}$ on *HE%* follows a similar trend as the dependency of $V_{app}$ on $Q_{ox}/Q_{dp}$ ratio (Figure 4a). *HE%* achieves the best values around the $V_{app}$ ≈ -0.1 V, at which the electrodeposited AuNPs possess a small particle size ($D_m$ ≈ 160 nm) with uniform distribution and a packing density ($N_D$) of ~3.2 particles/µm$^2$; at the same time they provide the best yield of electrochemical activity ($Q_{ox}/Q_{dp}$ ratio). Figure 4b reveals that *HE%* improves monotonously as the $C_{AuCl_4^-}$ decreases and reaches saturation at a concentration of 0.05 mM. On the other hand the $Q_{ox}/Q_{dp}$ ratio attains the best value at 0.1 mM. Hence, taking into account trends for both *HE%* and $Q_{ox}/Q_{dp}$ ratio, the following optimized conditions were chosen for Au ECD for the fabrication of the sensor: $V_{app}$ = -0.1 V and $C_{AuCl_4^-}$ = 0.1 mM, with MoS$_2$ loading of 50 µg.

**DNA–miRNA hybridization strategy.** The optimization strategy for the hybridization of **CP** with **TR** and **AP** is illustrated in the SI (Section S10 and Figure S11). The following terminology has been employed (SI, Table S3): (1) **CP** => **TR** => **AP** involves 3-steps: **CP**-immobilization followed by **CP–TR** hybridization and lastly by **TR–AP** hybridization; (2) **CP** => (**TR** + **AP**) involves 2-steps: **CP**-immobilization followed by the simultaneous hybridization of **CP** with **TR** and **AP**; and (3) **CP** => (**TR–AP**) involves 2-steps: **CP**-immobilization followed by hybridization of **CP** with previously hybridized **TR** and **AP** targets (**TR–AP**). It is clear from Figure S11 that the protocol **P#3**, **CP** => (**TR–AP**), yields the best hybridization efficiency. The *HE%* value improves by almost 30% compared to the sequential protocol (**P#1**).

**Importance of signal amplified probe (AP).** To verify the augmented function of the signal amplified probe (**AP**) in our proposed detection strategy, a miRNA-21 detection test was performed employing two different capture probes **CP** and **f-CP** (SI, Section S11 and Figure S12), with a target concentration of 10 fM. The **f-CP** probe is fully complementary to the target **TR** (Table S4) whereas **CP** is complimentary to a part of **TR** only, the remaining of **TR** is complementary to **AP**. Evidently, the 1-step hybridization of **f-CP** with **TR** (**AP#0**) cannot achieve the same hybridization efficiency, achieved by the 2-steps-protocol (**CP–TR** hybridization followed by **TR–AP** hybridization, **CP** => **TR** => **AP**). The involvement of **AP** can improve the *HE%* from 7% (for **AP#0**) to 10% (for **AP#3**) at the AuNPs@MoS$_2$/SPGE

sensor. Moreover, adopting the best hybridization protocol, that involves hybridization of immobilized **CP** with previously hybridized **TR** and **AP** targets, **CP** => (**TR**−**AP**), as described earlier in (Section S10, Figure S11), the efficiency can be improved further to ≈17% (**AP#4**) confirming the effectiveness of the amplification probe. Furthermore, both AuNPs@MoS$_2$/SPGE and AuNPs@SPGE sensors, exhibit a similar trend (Figure S12), which supports enhanced impact of the signal amplified probe (**AP**).

**Effect of CP concentration and hybridization time.** Further optimization studies conclude that the response signal of RuHex-assisted detection could be improved by: (1) lowering concentration of **CP** ($C_{CP}$) and (2) optimising the time ($T_H$) for the hybridization of **CP** with (**TR**−**AP**) targets. The best values for $C_{CP}$ are in the range of 0.3 ~ 0.1 µM. It is well known that the excessive probe DNA density ($\Gamma_{CP}$) would generate greater steric hindrance and reduce the hybridization efficiency (**HE%**). Similarly, the **HE%** exhibits significant improvement with the initial increase in $T_H$. However at much longer $T_H$ (> 45 mins), the **HE%** value becomes insensitive to the $T_H$ value.

## Chronocoulometric Detection of miRNA-21.

The chronocoulometric (CC) detection of miRNA-21 at the AuNPs@MoS$_2$/SPGE sensor was performed as a function of target concentration ($C_{TR}$) employing the $\Delta Q = [Q_{CP\text{-}TR\text{-}AP} - Q_{CP}]$ and the **HE%** = $[(\Delta Q \times 100) / Q_{CP}]$ as sensing parameters (Figure 7).

Two different MoS$_2$ NSs products (Figure 7) are compared for the fabrication of the sensor: MoS$_2$(1k) which consists of large and thick MoS$_2$ platelets (AuNPs@*MoS$_2$(1k)*/SPGE) and MoS$_2$(10k) consisting of thin and small nanosheets (AuNPs@*MoS$_2$(10k)*/SPGE). A representative AuNPs-coated bare SPGE (AuNPs@SPGE) was also employed for comparison purposes.

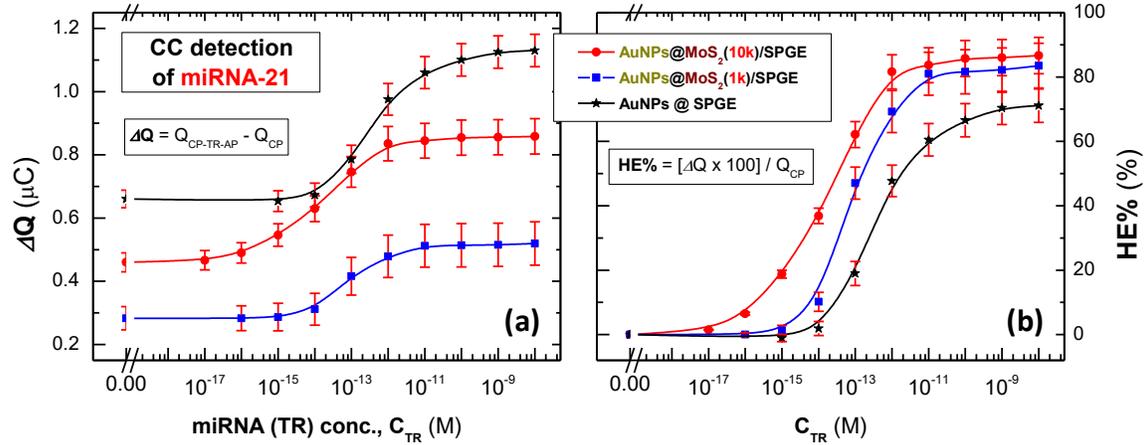

**Figure 7. Chronocoulometric (CC) detection of miRNA-21.** Logarithmic plot for (a) CC signal (*ΔQ*) and (b) corresponding hybridization efficiency (*HE%*) versus target miRNA (**TR**) concentration (*C$_{TR}$*) for the **CP**-immobilized on AuNPs (AuNPs@SPGE) and AuNPs@MoS$_2$ modified SPGEs employing MoS$_2$ nanosheets produced at centrifugation speeds of 1 k (AuNPs@*MoS$_2$(1k)*/SPGE) and 10 k rpm (AuNPs@*MoS$_2$(10k)*/SPGE).

*ΔQ* displays identical trends with **TR** concentration for all three sensors, exhibiting an initial rise and finally a plateau (Figure 7a). In the absence of **TR** (at *C$_{TR}$* = 0 M), relatively larger values of *ΔQ* are recorded at the AuNPs@SPGE sensor due to higher surface density of **CP** (*Γ$_{CP}$*). In contrast, both AuNPs@*MoS$_2$(10k)*/SPGE and AuNPs@*MoS$_2$(1k)*/SPGE sensors exhibit lower $Q_{CP}$ (at $C_{TR}$ = 0 M) because of a lower $\Gamma_{CP}$. The misleading underperformance of the AuNPs@MoS$_2$/SPGE, originating from the difference in the initial $Q_{CP}$ values (at $C_{TR}$ = 0 M), can be avoided by employing the *HE%* parameter (Figure 7b).

Actually, the AuNPs@MoS$_2$/SPGE sensor exhibits enhanced *HE%* values compared to the AuNPs@SPGE sensor. Interestingly, the AuNPs@*MoS$_2$(10k)*/SPGE sensor possessing the more electroactive MoS$_2$(10k) NSs exhibits the best *HE%* (≈88%) followed by the AuNPs@*MoS$_2$(1k)*/SPGE, while the AuNPs@SPGE sensor can only achieve a *HE%* of 70%.

The linear regimes of the *HE%* (also, *ΔQ*) versus the logarithm of $C_{TR}$ for the AuNPs@*MoS$_2$(10k)*/SPGE, AuNPs@*MoS$_2$(1k)*/SPGE, and AuNPs@SPGE sensors are estimated as [100 aM ~ 1 pM], [1 fM ~ 10 pM] and [10 fM ~ 10 pM], respectively. The "sensitivity" values are estimated from the linear fitting of these "linear regimes" as : 0.161 ± 0.007, 0.203 ± 0.012 and 0.203 ± 0.016 μC/log(M), respectively.

Interestingly, the AuNPs@*MoS$_2$(10k)*/SPGE sensor exhibits the best "experimental" limit of detection (*LoD*), of ~100 aM, followed by the AuNPs@*MoS$_2$(1k)*/SPGE, (*LoD* ≈ 1 fM) and AuNPs@SPGE (*LoD* ≈ 10 fM) sensors.

During the miRNA-21 detection study, the measurement error was estimated from the standard deviation of at least three independent experiments (n ⩾ 3), at every concentration of the miRNA target ($C_{TR}$). The relative standard deviation (RSD), obtained for the AuNPs@MoS$_2$/SPGE sensors, was 8.2%, slightly higher than that of AuNPs@SPGE sensor (6.5%). The lager RSD value is believed to originate from variations in coating surface areas associated with the drop-casting process of MoS$_2$ NSs.

**Selectivity of AuNPs@MoS$_2$/SPGE miRNA Sensor.** To evaluate the specificity of the as-proposed AuNPs@MoS$_2$/SPGE sensor, the interference from non-complementary target such as miRNA-155, as well as from base mismatched strands with the same concentration (10 fM) as that of the target (miRNA-21 = **T1**) were investigated (Table S5). The study, presented in Figure 8 (SI, Section S12), clearly reveals that all 3 sensors (utilizing MoS$_2$(10k), MoS$_2$(1k), and no MoS$_2$ NSs) become highly sensitive in the presence of complementary target **T1**. All the sensors can also sense the single-base mismatched target (**T2**), nevertheless at considerably lower signal. The *HE%*, measured at the AuNPs@*M-10k*/SPGE, for **T2** is only ~27% of that for **T1**, utilizing concentrations in the femtomolar range.

Interestingly, in the presence of either three-base mismatched (**T3**) or non-complementary target (miRNA-155: **T4**), no measurable signal was observed, which is a clear indication of an excellent sequence specificity of the proposed miRNA-sensor.

In the final experiment, performed in a complex medium with a mixture of all the targets (**Mix-T** = **T1**+**T2**+**T3**+**T4**, each target of 10 fM concentration), the AuNPs@*MoS$_2$(10k)*/SPGE sensor exhibits the best *HE%* (≈26.8%) followed by the AuNPs@*MoS$_2$(1k)*/SPGE (*HE%* ≈11.4%), while the AuNPs@SPGE sensor can only achieve a *HE%* of ≈2.4% (Figure 8).

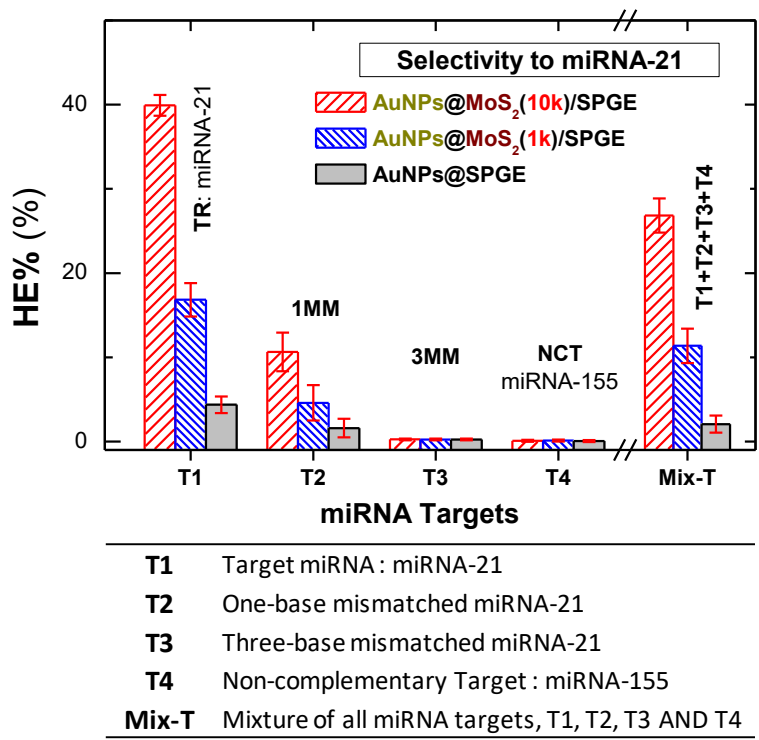

**Figure 8. Selectivity of AuNPS@MoS$_2$/SPGE sensor:** Hybridization Efficiency, *HE%*, measured at the **CP**-immobilized AuNPs@MoS$_2$/SPGE and AuNPs@SPGE sensors, in the presence of: miRNA-21 (**T1: TR**); Single-base mismatched strand (**T2: 1MM**); Three-base mismatched strand (**T3: 3MM**); and the interfering non-complementary target (**NCT**) (**T4: miRNA-155**), with the same concentration of 10 fM. **CP** is the anti-miRNA-21. Error bars represent the standard deviations estimated from at least three independent measurements.

- **CONCLUSIONS**

A new simple and sensitive electrochemical platform based on AuNPs@MoS$_2$ hybrid nanosheets coated on commercial disposable gold screen-printed electrode (SPGEs) has been developed for the detection of miRNA-21 using a chronocoulometric (CC) approach. The work consists of two major strands: (i) the controlled synthesis and tuning of AuNPs on MoS$_2$ NSs via CA electrochemical deposition (ECD); and (ii) the design of a simple new bioassay involving a label free signaling amplification probe for the chronocoulometric quantification of miRNA biomarker employing the AuNPs@MoS$_2$ platform.

Control on AuNP density and size was achieved, by regulating the kinetics of nucleation and growth through tuning of deposition-potential ($V_{app}$) and Au-precursor concentration ($C_{AuCl_4^-}$). By a combination of a statistical morphological and an electrochemically activity analysis, almost monodispersed AuNPs with small size (< 90 nm) and appropriate interparticle spacing were easily accomplished on the MoS$_2$ NSs. The CA Au-ECD method preserved the crystalline quality of the MoS$_2$ NSs and induced a p-type doping.

Following the AuNP optimization study, a detailed parametric CC study was undertaken to optimize each immobilization step. Our AuNPs@MoS$_2$/SPGEs sensor not only improved the **LoD** by 2 orders (≈100 aM) but also enhanced the **HE%** to ≈88%, when compared to bare AuNPs@SPGEs (**LoD** ≈ 10 fM, **HE%** ≈ 70%). Interestingly, the role of thin and small MoS$_2$ NSs was elucidated by demonstrating better sensing performance than that of thicker and larger counterparts. This work forms the first detailed and systematic study on sensitive CC detection of miRNA employing AuNPs/MoS$_2$ hybrids. The detection sensitivity is comparable to that obtained from systems based on complicated time consuming labelled amplification techniques. Here our design is based only on a simple non-labeled signaling probe (**AP**), which is cheap and easy to operate, avoiding complicated fabrication steps. The low detection limit originates from the controlled packing density of **CP**s, achieved by their self-assembly on AuNPs, and the intimate coupling between AuNPs and $MoS_2$. Our methodology provides important guidelines for the sensitive detection of miRNA cancer diagnostics.

- **EXPERIMENTAL SECTION**

**Synthesis of $MoS_2$ Nanosheets.** $MoS_2$ NSs were synthesized by the grinding ionic liquid assisted exfoliation method followed by size selection ultra-centrifugation steps as reported in our earlier publication.[7] $MoS_2$ NSs pelleted at 1000 rpm and 10,000 rpm are abbreviated as $MoS_2$(1k) and $MoS_2$(10k) respectively. $MoS_2$ inks were prepared by dispersing 5 mg of $MoS_2$ NSs in 1 ml DMF and 50 μl of Nafion solution under adequate ultrasonication.

**Chronocoulometry Detection of miRNA-21.** The chronocoulometry (CC) is used here as the main technique for the detection of miRNA-21, by quantifying the saturated amount of charge compensated RuHex redox marker at the hybridized electrode system (Step 6 in Figure 1). The overall principle of CC DNA detection is based on determination of surface-confined redox species like $[Ru(NH_3)_6]^{3+}$ (RuHex) at the DNA-electrode system,[1, 4, 6, 34-35, 41-42] where the cationic redox markers, RuHex, can electrostatically interact with the negative phosphate groups of the DNA or RNA. The numerical analysis of the CC data was performed through Anson plots as demonstrated in the SI (Figure S1) and described in Section S2. In a typical CC experiment the following steps were followed. At first CC was performed in blank TE buffer. From the Anson Plot (CC-plot), which provides the measured charge ($Q$) versus the square-root of time ($t^{1/2}$), the double-layer charge term ($Q_{dl}$) was estimated from the y-axis-intercept ($Q_{blank} = Q_{dl}$). Next, the surface-confined redox marker, RuHex, was introduced to the TE buffer solution at a concentration that provided saturation with the probe DNA layer. The adsorption isotherm of RuHex was investigated to optimize the RuHex concentration ($C_{RuHex}$) and the CC-conditions. In the presence of RuHex, the y-axis-intercept of the CC-plot gave the $Q_{RuHex}$ value. The value of surface excess of RuHex, $Q_{ad} = (Q_{RuHex} - Q_{blank}) = nFA\Gamma_0$, was calculated from the difference in the CC intercepts in the absence ($Q_{blank}$) and presence ($Q_{RuHex}$) of RuHex.

For the miRNA-detection performance study, the concentration of miRNA-21 targets ($C_{TR}$) was varied. For CC-based miRNA detection, the change in signal, $\Delta Q$ (= ($Q_{CP-TR-AP} - Q_{CP}$)) and the corresponding hybridization efficiency, $HE\%$ (= ($\Delta Q \times 100$) / $Q_{CP}$) were treated as the sensing parameters: where, $Q_{CP}$ and $Q_{CP-TR-AP}$ represent the $Q_{ad}$ values measured after **CP** immobilization and after its hybridization with **TR** and **AP**, respectively.

- **ASSOCIATED CONTENT**

**Supporting Information**

Experimental section (materials, methods and instruments, electrochemical sensing methods, synthesis of MoS$_2$ nanosheets, synthesis of AuNPs@MoS$_2$ hybrids on SPGEs, and assembly of **CP–TR–AP** on AuNPs@MoS$_2$/SPGE); Numerical analysis of chronocoulometric data; Effects of pretreatment of MoS$_2$/SPGEs and electro-deposition methods on gold nanoparticle dispersion; Equations used for the electrochemical characterizations of AuNPs and Fe(CN)$_6^{3-/4-}$ redox activity; Confirmation of sensor-fabrication strategy, via the redox characteristics of Methylene Blue (MB) detection probe; Equation deployed for the adsorption isotherm of RuHex and effect of $C_{RuHex}$ on the voltammetric responses of RuHex; Optimization of **CP–TR–AP** hybridization strategy; Control experiment to establish the amplified function of **AP** probe; Chronocoulometric responses on **CP–TR** hybridization in absence of RuHex; and Tables presenting the sequences of oligonucleotides used in this study.

The following files are available free of charge.

- **AUTHOR INFORMATION**

**Corresponding Author**

*P. Papakonstantinou. E-mail: p.papakonstantinou@ulster.ac.uk.

**Funding Sources**

**Notes**

The authors declare no competing financial interest.

- **ACKNOWLEDGMENTS**

The authors acknowledge support from British Council (Newton fund, Institutional Links, Ref: 216182787).

# Graphical Abstract or TOC Graphic

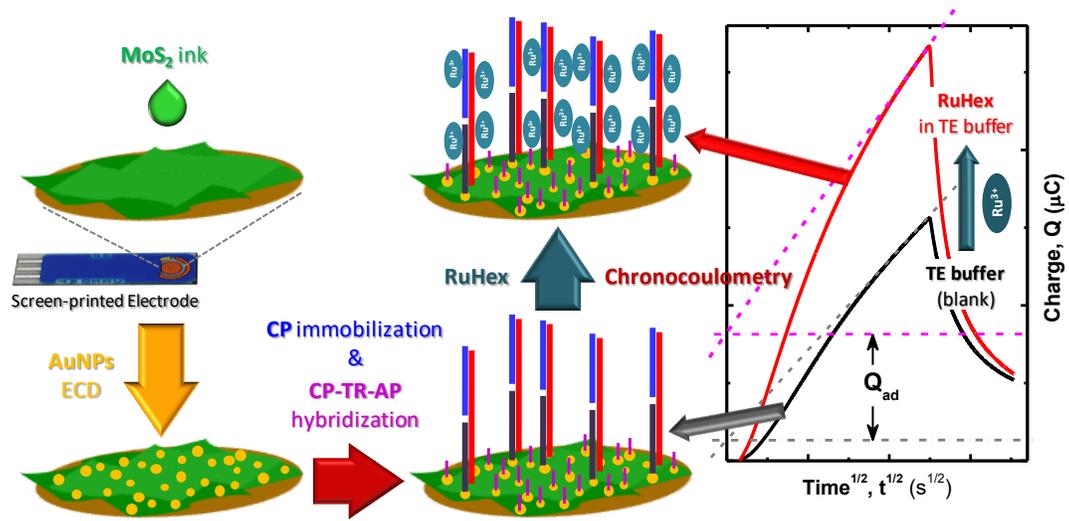

# Supporting Information

Sensitive Chronocoulometric Detection of miRNA at Screen-printed Electrodes modified by Gold decorated MoS$_2$ Nanosheets

*Abhijit Ganguly,[†] John Benson,[‡] and Pagona Papakonstantinou[\*,†]*

[†]School of Engineering, Engineering Research Institute, Ulster University, Newtownabbey BT37 0QB, United Kingdom

[‡]2-DTech, Core Technology Facility, 46 Grafton St., Manchester M13 9NT, United Kingdom

[\*]Corresponding author, e mail address: *p.papakonstantinou@ulster.ac.uk*

**List of Contents**

## S1. Experimental Section.

**Materials.** Molybdenum (IV) sulphide ($MoS_2$) powder (<2 μm, 99.0%), Gold(III) chloride hydrate ($HAuCl_4 \cdot xH_2O$, 99.999% metal basis), Tris (2-carboxyethyl) phosphine hydrochloride (TCEP), 6-Mercapto-1-Hexanol (MCH, 97%), and Hexaammineruthenium(III) chloride ($[Ru(NH_3)_6]^{3+}$, RuHex) were purchased from Sigma-Aldrich. Solvents/media like N,N-dimethylformamide (DMF, ≥99.9%), Acetone (≥99.8%), room temperature ionic liquid (RTIL), 1-butyl-3-methylimidazolium hexafluorophosphate ($BMIMPF_6$, ≥97.0%), and TE buffer solution (10mM Tris–HCl, 1mM EDTA, pH 7.4) were also supplied from Sigma-Aldrich.

The Thiol-modified capture ssDNA probe (**CP**: anti-miRNA-21), the target miRNA-21 (**TR**), and the signal amplified ssDNA probe (**AP**) were purchased from Ella Biotec (Table S1). Ultrapure water (resistivity of 18.2 MΩ·cm, Millipore) was used to prepare all aqueous solutions.

**Table S1:** Sequences of Oligonucleotides used

|    | **Oligonucleotides used**           | **Sequence** (from 5′ to 3′)          |
|----|-------------------------------------|---------------------------------------|
| **CP** | Thiol-modified Capture Probe DNA | 5′-Thiol- TCA ACA TCA GT -3′          |
| **TR** | Target RNA : miRNA-21            | 5′- UAG CUU AUC AGA CUG AUG UUG A -3′ |
| **AP** | Signal Amplified Probe DNA       | 5′- CTG ATA AGC TA -3′                |

Immobilization buffer (**I-buffer**): 10 mM Tris-HCl + 1 mM EDTA + 0.1 M NaCl + 10 mM TCEP (pH 7.4). Electrochemistry buffer (**E-buffer**): 10 mM Tris-HCl (pH 7.4). Washing buffer (**W-buffer**): 10 mM Tris-HCl (pH 7.4). Hybridization buffer (**H-buffer**): 10 mM Phosphate buffer + 0.25 M NaCl (pH 7.4). MCH solution: 1 mM 6-mercapto-1-hexanol (MCH) in water. RuHex solution: 10 mM Hexammineruthenium(III) chloride $[Ru(NH_3)_6]^{3+}$ in water.

Mechanical grinding of $MoS_2$ platelets was performed by an agate mortar and pestle grinder system (RM200, Retsch GmbH) and the size selection sequential centrifugation steps by a Thermo Scientific Sorvall ST-16 Centrifuge system. Au decorated $MoS_2$ based electrodes were fabricated using commercial disposable screen printed gold electrodes (SPGEs) from DropSens (DS-C220AT) as a supporting electrode. SPGEs were built with a gold working electrode (Au-WE) of 4 mm diameter, a gold counter electrode (CE) and a silver pseudo reference electrode (RE).

**Methods and Instruments.** Surface morphology of all the samples were studied using a Scanning electron microscope (SEM, FEI Quanta 200 2D) at an accelerating voltage of 15kV. For SEM observations, $MoS_2$ coated SPGEs (**$MoS_2$/SPGE**) or Au electrodeposited (**AuNPs@$MoS_2$/SPGE**) were loaded directly inside the SEM-chamber.

X-ray photoelectron spectroscopy (XPS) analysis was carried out using a Kratos AXIS ultra DLD with an Al Kα (hν = 1486.6eV) x-ray source. Elemental quantification was performed after Shirley background correction of all the spectra and calibration of the binding energies with respect to the C 1s line at 284.5 ± 0.2 eV associated with graphitic carbon.

Electrochemical deposition (ECD) of AuNPs and electrochemical characterizations were performed by an Autolab, PGSTAT/FRA system, on the as-purchased bare SPGEs, or fabricated $MoS_2$/SPGE, and AuNP@$MoS_2$/SPGE. All washing and electrochemical sensing steps were performed in TE buffer solution (pH 7.4). For all ECD processes and characterizations, a 40 μl of electrolyte was employed, in order to cover the whole sensor-zone including the counter and reference electrode areas.

**Electrochemical Sensing Methods.** All electrodes, before and after probe immobilization and/or hybridization steps were characterized by Cyclic (CV), Linear Sweep LSV) and Differential Pulse Voltammetry (DPV). If not specified, CVs were recorded within the potential range from −0.75 V to +0.2 V, at a scan-rate of 50 mV/s. LSVs were carried out within +0.5 V to −0.8 V potential range, at a scan-rate of 5 mV/s. DPVs were performed within the potential range from −0.75 V to +0.2 V under a pulse amplitude of 50 mV, pulse width of 0.05 s, pulse period of 0.2 s and an increasing potential of 4 mV.

For the electrochemical detection of miRNA-21, Chronocoulometry (CC) was conducted at a pulse period of 250 ms to quantify capture probe (**CP**) DNA surface density and monitor miRNA hybridization. For CC measurements at AuNPs@MoS$_2$/SPGEs, the pulse width was extended to 800 mV (applied potential from +0.2 V to -0.6 V), employing 14.5 μM of RuHex to completely neutralize the negative charges of **CP** for efficient quantification. Accordingly, at the bare SPGEs or AuNPs@SPGEs, the pulse width was kept at 700 mV (applied potential from +0.2 V to -0.5 V), using 50 μM of RuHex.

**Synthesis of MoS$_2$ Nanosheets.** The exfoliation method for the synthesis of MoS$_2$ Nanosheets (MoS$_2$ NS) involves the mechanical grinding of MoS$_2$ platelets in a minute quantity of room temperature ionic liquid (RTIL) to produce a gel. During grinding, the RTIL protects every newly exposed MoS$_2$ surface by adsorbing onto the surface, keeping the sheets separated and avoiding restacking. Following grinding, the resulting gel is subjected to multiple washing steps in a mixture of DMF and acetone, to remove the RTIL and is isolated using centrifugation at high speed (10,000 rpm). Finally, the clean ground product, consisting of an assortment of sheets of various sizes and thicknesses, is dispersed in pure DMF (via short sonication time) and subsequently is subjected to sequential centrifugation steps involving progressively increasing centrifugation speeds from 500 to 10,000 rpm, to isolate different grades of MoS$_2$ NSs. As we have demonstrated in a series of previous papers the sequential centrifugation of the supernatant at progressively higher centrifugation speeds allows the isolation of thinner and smaller nanosheets.[1-4] Large and thick platelets are pelleted at low speeds and small durations, generating high yield. However, the yield of smaller and thinner nanosheets is smaller and requires longer centrifugation times at high speeds. The nanosheets are free of defects in the basal plane and retain the semiconducting phase (2H) of the bulk starting material. MoS$_2$ NSs pelleted at 1000 rpm and 10,000 rpm are abbreviated as MoS$_2$(1k) and MoS$_2$(10k) respectively. MoS$_2$ inks were prepared by dispersing 5 mg of MoS$_2$ NSs in 1 ml DMF and 50 μl of Nafion solution under adequate ultrasonication.

**Synthesis of AuNPs@MoS$_2$ Hybrids on SPGEs.** Prior to the sensor preparation, the gold working electrode of the commercial SPGE (Au-WE) was pretreated electrochemically in 0.5 M H$_2$SO$_4$ aqueous solution by potential cycling (CV) in a range of -0.3 to +1.5 V at a scan-rate of 100 mV/s until the CV characteristic of a clean Au electrode was obtained. Then, the Au electrode was washed thoroughly with redistilled deionized water and dried under nitrogen gas. The actual working electrode (WE) was fabricated by drop-drying the desired volume of the MoS$_2$ ink onto the Au-WE (Figure 1, Step 1). For a typical ECD process, a 10 μl of MoS$_2$ ink (5 mg/ml) was used to attain the desired catalyst-loading of 50 μg (≈398 μg/cm$^2$). Preceding Au electrodeposition, the working electrode (MoS$_2$/SPGE) was pretreated electrochemically in aqueous electrolyte (1x PBS or 0.5 M H$_2$SO$_4$ solution) via cyclic voltammetry (CV), in the potential range of -0.2 to +0.9 V at a scan-rate of 100 mV/s, until the voltammograms became stable and repetitive; followed by thorough washing with redistilled deionized water. Subsequently, AuNPs were deposited onto MoS$_2$/SPGE using either static chronoamperometry (CA) or dynamic cyclic voltammetry (CV) methods, in 0.5 M H$_2$SO$_4$ solution containing HAuCl$_4$ gold precursor (Step 2 in Figure 1). For all ECD processes and characterizations, a 40 μl of electrolyte was dropped on the WE of SPGE, in order to cover the whole sensor-zone.

**Assembly of CP–TR–AP on AuNPs@MoS$_2$/SPGE.** Following the Au-electrodeposition (Au-ECD), the AuNPs@MoS$_2$/SPGE working electrode was pretreated electrochemically in 0.5 M H$_2$SO$_4$ aqueous solution by potential cycling (CV) in the potential range of -0.2 to +0.9 V at a scan-rate of 100 mV/s until the cyclic voltammograms became stable; this pretreatment served to clean and activate the WE part.[5-6] Next, the electrode was washed carefully with 1x PBS solution and dried in ambient. Subsequently, a 10 μL of probe solution, consisting of **CP** (anti-miRNA-21, 1 μM) with 10 mM TCEP, was dropped on the cleaned AuNPs@MoS$_2$/SPGE working area, and incubated for 18 h at 4 °C (Figure 1, Step 3). Following the self-assembly of **CP**, the non-specific binding areas were passivated with 1 mM MCH (10 μL) for 1 h at room temperature (Figure 1, Step 4). At the final step, prior to the electrochemical sensing or characterizations, the **CP**-MCH functionalized AuNPs@MoS$_2$/SPGE was cleaned with 1x PBS several times and stored at 4 °C. The hybridization procedure was performed by dropping the target miRNA-21 (**TR**, 5 μL, predefined concentration) and the signal amplified **AP** probe (5 μL, predefined concentration) onto the **CP**-MCH-AuNPs@MoS$_2$/SPGE working area and incubating for 30~120 min at room temperature (Figure 1, Step 5). Lastly, the hybridized electrode was rinsed with 1x PBS solution to remove non-specifically adsorbed **TR** or **AP**.

## S2. Numerical Analysis of Chronocoulometry Data.

The numerical analysis has been performed based on the Anson plots demonstrated in Figure S1:

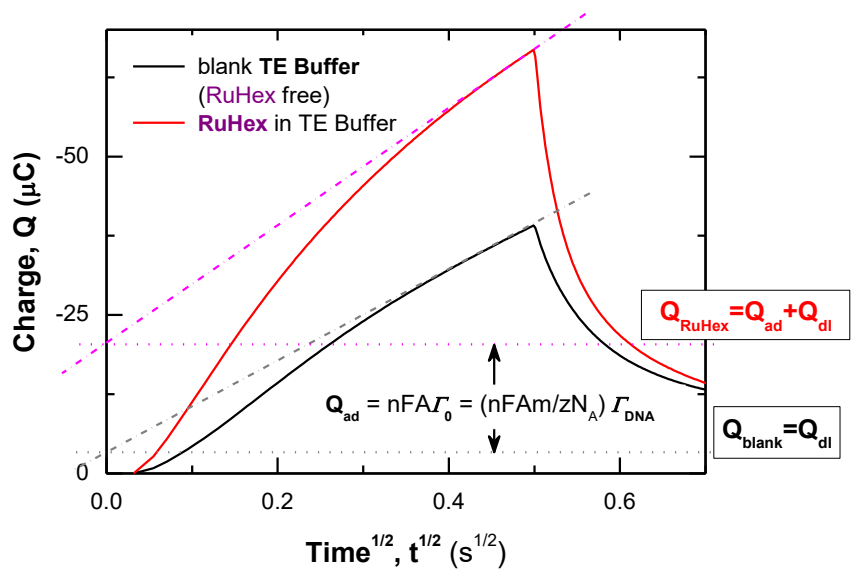

**Figure S1. Conversion of chronocoulometry data to Anson plot.** Example of a typical Anson plot exhibiting the analysis method for the estimation of DNA coverage on AuNPs@MoS$_2$/SPGEs: $Q_{dl}$ = double-layer charge at the electrode/electrolyte interface. $Q_{ad}$ = charge of surface-adsorbed RuHex. $\Gamma_{DNA} = \Gamma_0(z/m)N_A$ = surface-coverage of DNA.

The total measured charge $Q$ follows the integrated Cottrell expression as a function of time $t$:

$$Q = (2nFAD_0^{1/2}C_0)\cdot(t/\pi)^{1/2} + Q_{dl} + Q_{ad} \tag{1}$$

From Faraday's law,

$$Q_{ad} = (Q_{RuHex} - Q_{blank}) = nFA\Gamma_0 \tag{2}$$

Here, n: number of electrons per molecule for reduction = 1 (in this case); F: Faraday constant (C/equiv); A: electrode area = 0.126 cm² (for a WE of 4 mm diameter); $D_0$: diffusion coefficient (cm²/s), and $C_o$: bulk concentration (mol/cm²).

As shown in Fig. S1, the $Q_{dl}$ and $Q_{ad}$ were estimated from the CC intercepts at t = 0 in the absence and presence of RuHex, respectively.

The saturated surface excess of redox marker was converted to DNA surface coverage density by

$$\Gamma_{DNA} = \Gamma_0(z/m)N_A \qquad (3)$$

where m = number of bases in the DNA sequences = 11; z = charge of the RuHex = 3; and $N_A$ = Avogadro's number (/mol).

For chronocoulometric detection of **TR**, the *ΔQ* was treated as the sensing parameter:

$$\Delta Q = (Q_{CP\text{-}TR\text{-}AP} - Q_{CP}) \qquad (4)$$

Corresponding Hybridization Efficiency, **HE%** was estimated by:

$$\mathbf{HE\%} = (\Delta Q \times 100) / Q_{CP} \qquad (5)$$

where, $Q_{CP}$ and $Q_{CP\text{-}TR\text{-}AP}$ represent the $Q_{ad}$ values, measured at the AuNPs@MoS₂/SPGEs, after the capture DNA probe (**CP**) immobilization and after its hybridization with the target miRNA (**TR**), respectively.

## S3. Effect of Pre-treatment of MoS₂/SPGEs on Au–ECD.

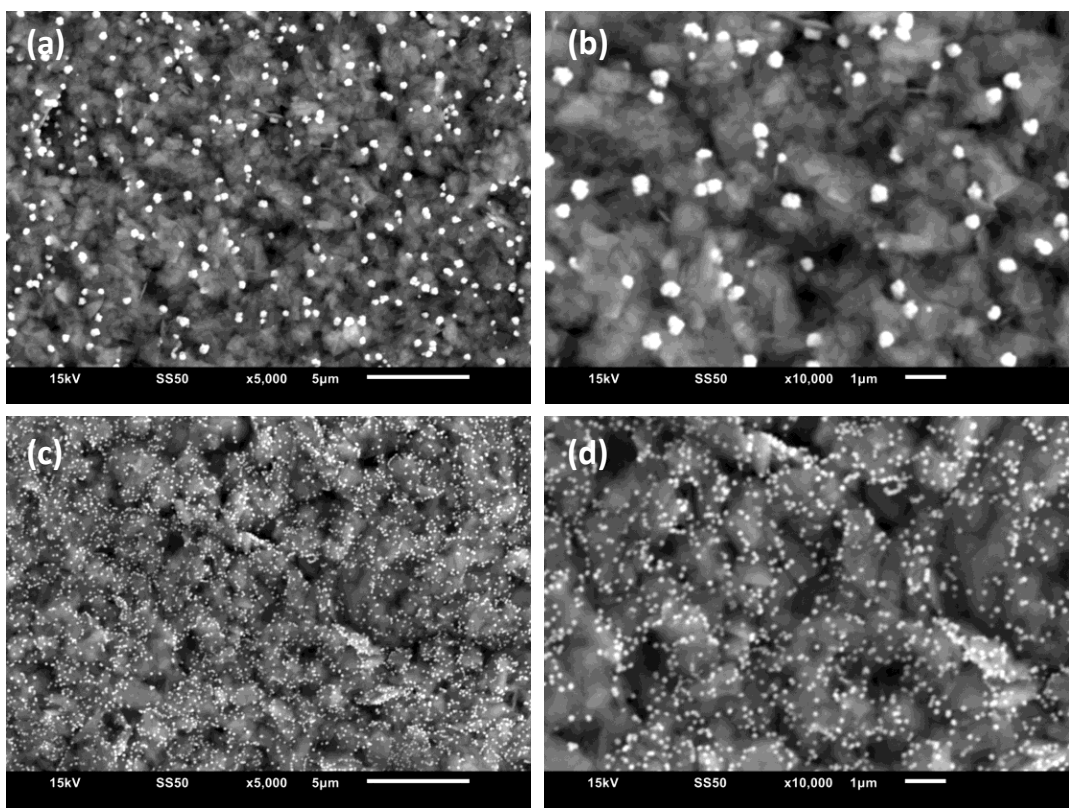

**Figure S2. Effect of pretreatment of MoS₂/SPGEs on Au-ECD.** SEM images of AuNPs electrodeposited via CV on the MoS₂/SPGEs, which are pre-treated in (a-b) 1x PBS and (c-d) 0.5 M H₂SO₄ aqueous solution respectively. Pre-treatment conditions: potential cycling (CV) from -0.2 to +0.9 V at a scan-rate of 100 mV/s for 10 cycles. AuNPs electrodeposition was performed via CV under potential scan-range: +0.9 to 0.0 V, at a scan-rate of 100 mV/s, for 10 cycles, in 1 mM HAuCl₄ in 0.5 M H₂SO₄. MoS₂ Loading: 50 µg.

Pretreatment of the MoS$_2$/SPGE working electrode was crucial for achieving reproducible NP deposition as it served the dual purpose of both cleaning of organic solvent residue (DMF) and also activating the electrode surface for subsequent AuNP formation.

Generally, in the case of biosensor-electrodes, aqueous solution of neutral pH, such as PBS or TE buffer (pH 7.4) is an obvious choice of electrolyte for such pre-treatment. Prior to Au electrodeposition, the working electrode (MoS$_2$/SPGE) was pretreated electrochemically in aqueous electrolyte by CV until the cyclic voltammograms became stable and repetitive. Interestingly, our studies revealed that electrochemical pre-treatment (CV) in PBS solution lead to a low density of large AuNPs (Figures S2a and S2b). On the contrary, when the MoS$_2$/SPGE working electrode was pretreated in 0.5 M H$_2$SO$_4$ (Figures S2c and S2d) a high density of small particle size was evident under the same ECD-conditions, demonstrating better control in Au-electrodeposition.

## S4. Comparison of Electro-deposition Methods: cyclic voltammetry vs. chromoamperometry.

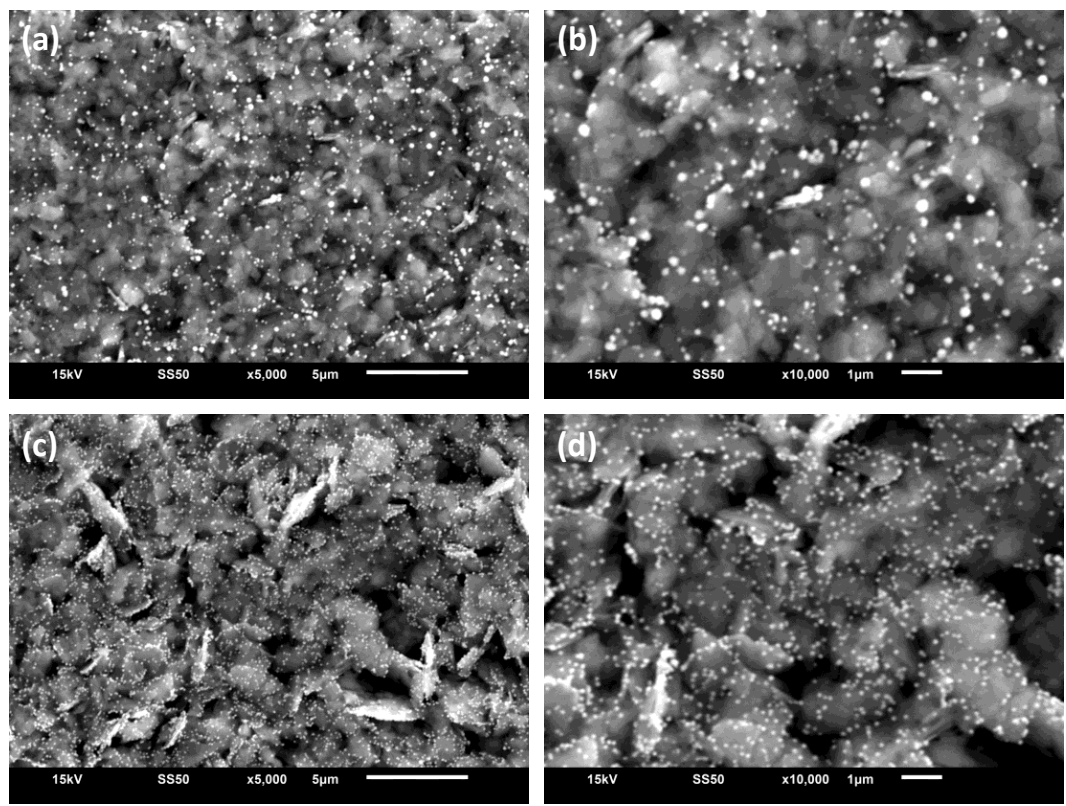

**Figure S3. Comparison of cyclic voltammetry (CV) and chronoamperometry (CA) electro-deposition methods.** SEM images of AuNPs electrodeposited on $MoS_2$/SPGEs via (a-b) cyclic voltammetry and (c-d) chronoamperometry routes, respectively. Electrolyte: 1 mM $HAuCl_4$ in 0.5 M $H_2SO_4$. $MoS_2$ Loading: 50 µg. **(a-b)** CV route: from +0.9 to 0.0 V, scan-rate: 100 mV/s, 10 cycles. **(c-d)** CA route: at 0.0 V, for 360 s. For both CV and CA, 0.0 V was chosen as the deposition potential, and the total duration of deposition was 6 min.

A literature survey identified that AuNPs can be formed by using almost all kinds of electrochemical techniques, which can be classified in two main categories: static and dynamic. Static methods include chronoamperometry (CA) or galvanostatic (GS) approaches by applying an explicit potential or current for a specific duration, whereas dynamic methods include cyclic voltammetry (CV) by scanning the potential in a specified range using appropriate scan-rate and number of scan-cycles.

It was vividly evident from our studies that the CV based ECD approach (Figures S3a and S3b) clearly suffers from coalescence phenomenona. On the contrary, the CA route can provide much better control in Au ECD, with homogeneous distribution of small AuNPs onto $MoS_2$/SPGE, as observed in Figures S3c and S3d. Similar observations have been reported earlier by Hezard et al.[5] for AuNPs-deposition on bare GC electrode.

**Effect of potential scan-rate during CV electrodeposition.** It was observed that, low potential scan-rates during CV-ECD could lead to a higher number density of AuNPs with smaller overall particle-size (Figure S4). It is presumed that at low scan-rates, the nucleation rate dominates over the growth rate, hence the number density of AuNPs increases, while the NPs-size deceases. However, even at a low scan-rate of 10 mV/s (Figure S4a), the CV-deposited AuNPs were inferior to the CA-deposited NPs in terms of both size and distribution. It was concluded that the electrodeposition process was better controlled under a static applied potential than that under dynamic potential scan.

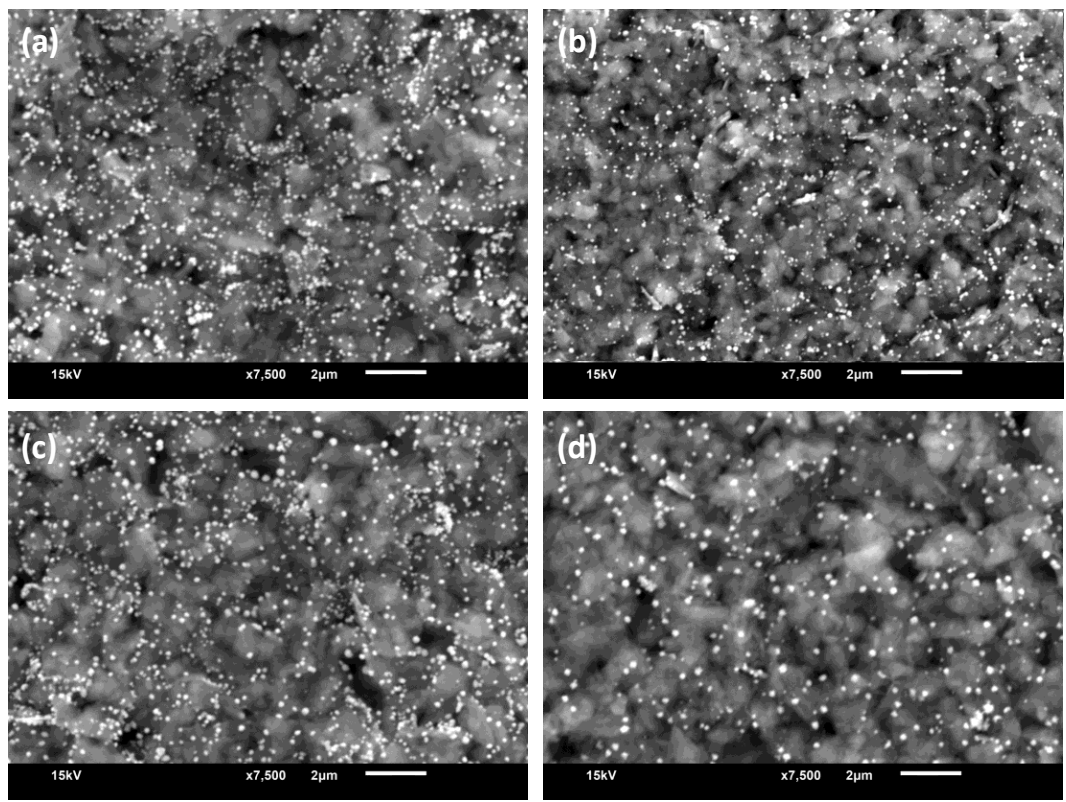

**Figure S4. Effect of potential scan-rate during CV electrodeposition of AuNPs on MoS$_2$ NSs.** SEM images of AuNPs electrodeposited on the MoS$_2$/SPGEs via cyclic voltammetry (CV) at different potential scan-rates: (a) 10, (b) 50, (c) 150, and (d) 250 mV/s. Potential scan-range: +0.9 to 0.0 V, scan-duration: ~6 min. Electrolyte: 1 mM HAuCl$_4$ in 0.5 M H$_2$SO$_4$. MoS$_2$ Loading: 50 μg.

## S5. Effect of Applied Potential $V_{app}$ = -0.3V.

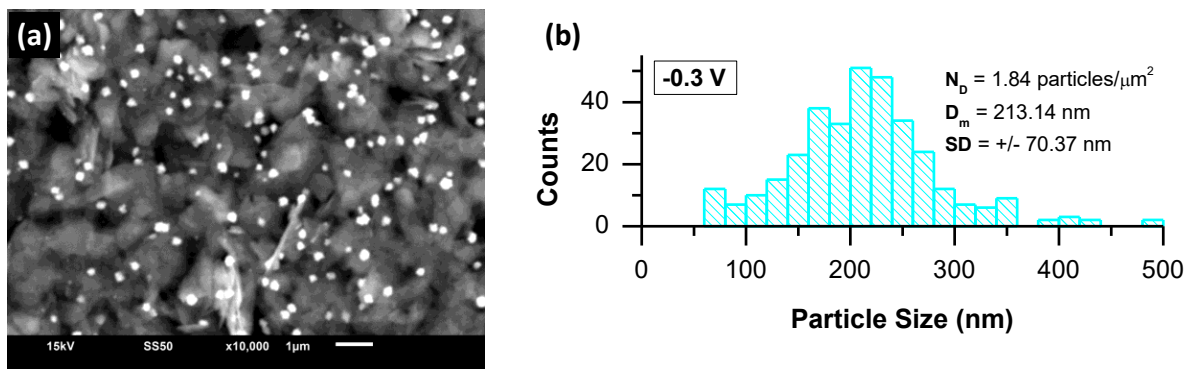

**Figure S5. Effect of applied potential $V_{app}$.** (a) SEM images of AuNPs electrodeposited via chronoamperometry (CA) on MoS$_2$/SPGEs at an applied potential, $V_{app}$ = -0.3 V. Potential scan-duration: 360 s. Electrolyte: 1 mM HAuCl$_4$ in 0.5 M H$_2$SO$_4$. MoS$_2$ Loading: 50 µg. (b) Statistical analysis: corresponding particle size histograms of AuNPs electrodeposited on the MoS$_2$/SPGEs. $N_D$: Particle Number Density, $D_m$: Mean Diameter of AuNP, and $SD$: Standard Deviation of NP-size.

## S6. Electrochemical Characterization of AuNPs@MoS$_2$ Hybrid.

**Electrochemical characterizations of AuNPs.**

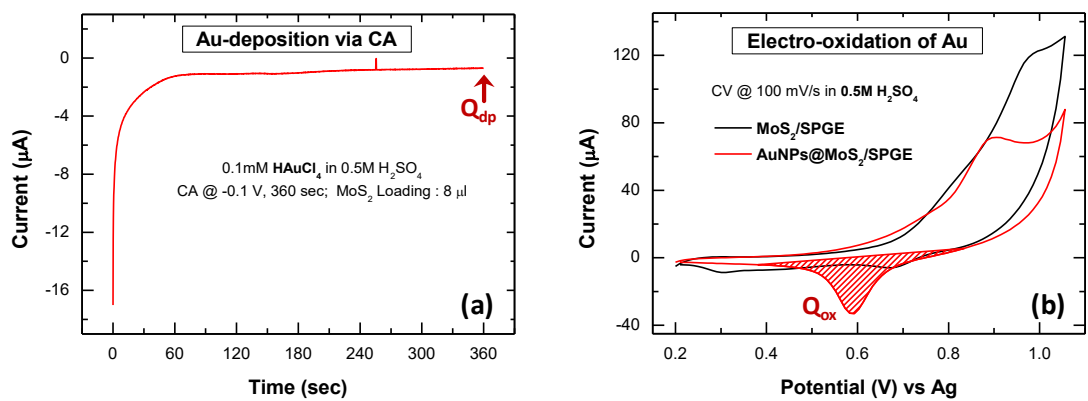

**Figure S6. Electrochemical characterisation of electrodeposited AuNPs on MoS$_2$ nanosheets (NSs).** (a) Representative current transients recorded at a MoS$_2$/SPGE electrode during Au electrodeposition via chronoamperometry (CA). (b) Representative CV spectra of pristine MoS$_2$/SPGE and AuNPs@MoS$_2$/SPGE, in 0.5 M H$_2$SO$_4$ at 100 mV/s vs. Ag pseudo RE.

Figure S6a represents typical current transients recorded at a MoS$_2$/SPGE (MoS$_2$ Loading: 50 µg) during Au electrodeposition via CA, in 0.1 mM HAuCl$_4$ in 0.5 M H$_2$SO$_4$.

The amount of charge involved for the AuNP formation, abbreviated as $Q_{dp}$ was estimated by integrating the corresponding current transient curves obtained by CA.[5, 7]

Figure S6b shows that the hybrids exhibit a typical characteristic voltammogram of gold electrodes in a 0.5 M H$_2$SO$_4$ solution, according to the following reaction:[5]

$$Au + H_2O \leftrightarrow AuO + 2H^+ + 2e^- \qquad (6)$$

Following the oxidation of AuNP-surface during the forward scan (manifested by an anodic peak around 0.83 V vs. Ag RE), the reduction of these oxides occurs during the backward scan as evidenced from a well-defined reduction peak at ca 0.49 V (Figure S6b).

The electrodeposited AuNPs were further characterized by evaluating the charges related to the reduction of Au-oxides, abbreviated as $\boldsymbol{Q_{ox}}$, by simply estimating the area under the corresponding reduction peak.[5, 7]

### Redox activity of Fe(CN)$_6^{3-/4-}$ at AuNPs@MoS$_2$/SPGE.

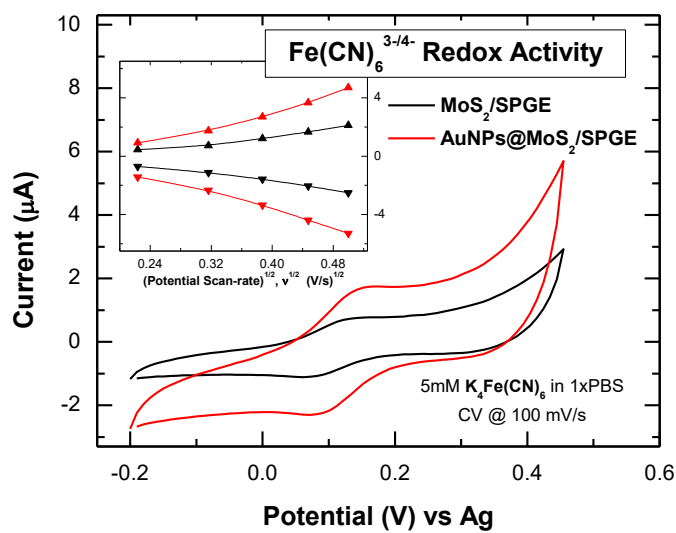

**Figure S7. Electrochemical Redox Activity.** Representative CV responses and (inset) effect of scan-rates on the redox peak currents corresponding to Fe(CN)$_6^{3-/4-}$ redox activities, for pristine MoS$_2$ and AuNPs@MoS$_2$ hybrids. $V_{app}$: -1.0 V, Potential scan-duration: 360 s. $\boldsymbol{C_{AuCl_4^-}}$: 0.1 mM HAuCl$_4$ in 0.5 M H$_2$SO$_4$. MoS$_2$ loading: 50 µg.

Electrochemical characterization on the redox activity of $Fe(CN)_6^{3-/4-}$ at the AuNPs@MoS$_2$/SPGE and MoS$_2$/SPGE revealed the higher electrocatalytic efficiency of AuNPs@MoS$_2$ hybrids over that of bare MoS$_2$ NSs. The well-defined anodic and cathodic peaks, produced by the oxidation and reduction of $Fe(CN)_6^{3-/4-}$ redox probe, are significantly enhanced following the AuNPs electrodeposition (Figure S8). Furthermore, the linear relationship of the redox peak currents ($I_p$) with the square root of scan-rates ($v^{1/2}$) of potential (inset of Figure S8), suggests diffusion-controlled mass transport, following the Randles-Sevcik equation:

$$I_p = (2.69 \times 10^5)\, n^{3/2}\, A\, D^{1/2}\, C\, v^{1/2} \qquad (7)$$

where **n** is the number of electrons participating in the redox reaction. **D** and **C** are the diffusion coefficient ($7.6 \times 10^{-6}$ cm$^2$/s) and the concentration (mol/cm$^3$) of K$_4$Fe(CN)$_6$ in solution. *A* represents the electroactive surface area of the electrode or ***ESA*** (cm$^2$).

Through the linear fitting, it was estimated that the ***ESA*** of AuNPs@MoS$_2$ hybrid NSs is almost ~4.4 times higher than that of pristine MoS$_2$ NSs. The improved performance provided by AuNPs is attributed to improved conductivity, enhanced electroactive sites for the redox activity and their intimate coupling with the MoS$_2$ NSs favoring efficient electron-transfer pathway from the redox-event to the gold SPE current-collector.

## S7. XPS study on AuNPs@MoS$_2$ Hybrids Nanosheets.

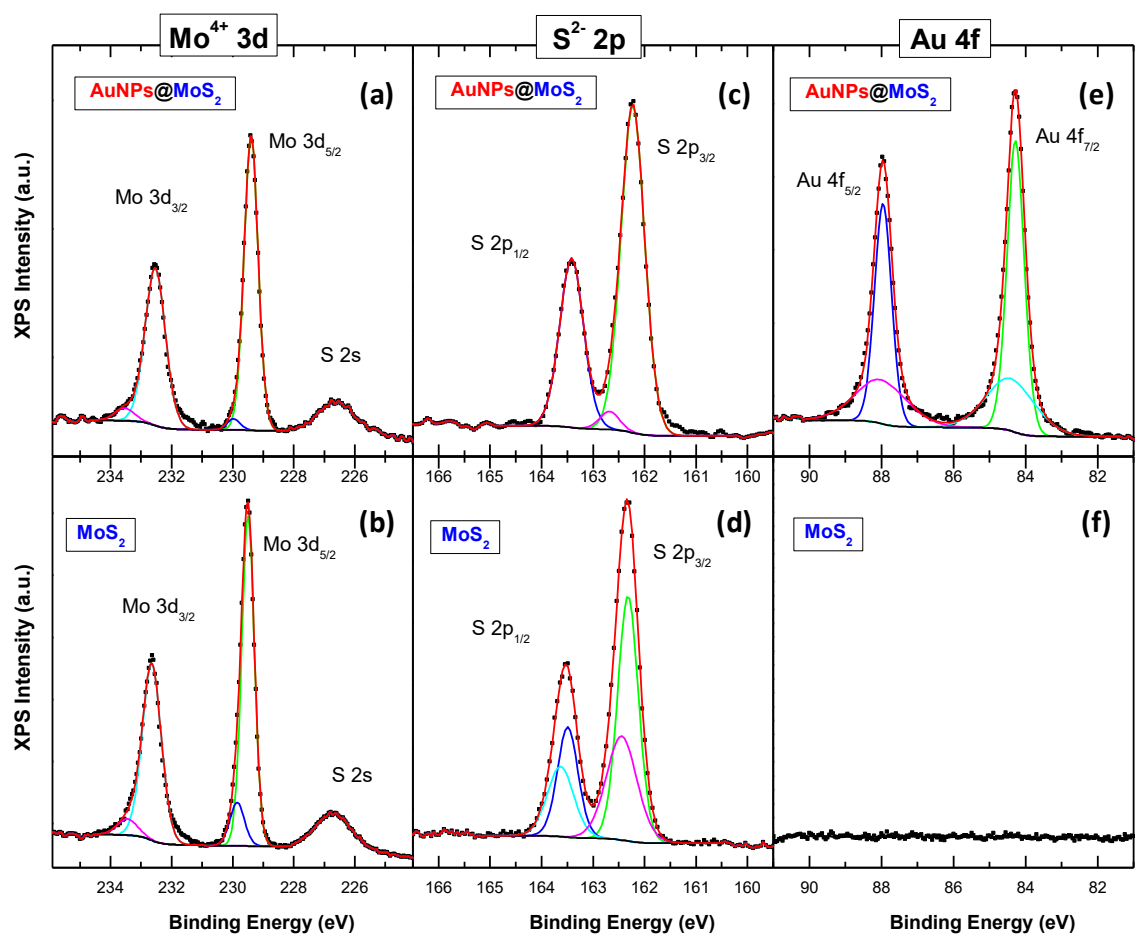

**Figure S8. Elemental characterization of AuNPs@MoS$_2$ hybrids NSs.** High-resolution XPS spectra of AuNPs@MoS$_2$ (top) and pristine MoS$_2$ NSs (bottom) drop-casted on SPGEs: (a-b) Mo$^{4+}$ 3d, (c-d) S$^{2-}$ 2p, and (e-f) Au 4f. All spectra are corrected by Shirley background and calibrated with reference to the C 1s line at 284.5 ± 0.2 eV associated with graphitic carbon. For AuNPs@MoS$_2$ hybrid NSs, the AuNPs are electrodeposited via chronoamperometry for 360 s: $V_{app}$: -1.0 V. $C_{AuCl_4^-}$: 0.1 mM HAuCl$_4$ in 0.5 M H$_2$SO$_4$. MoS$_2$ Loading: 50 μg.

# S8. Confirmation of Sensor-fabrication Strategy: Immobilization / Hybridization steps.

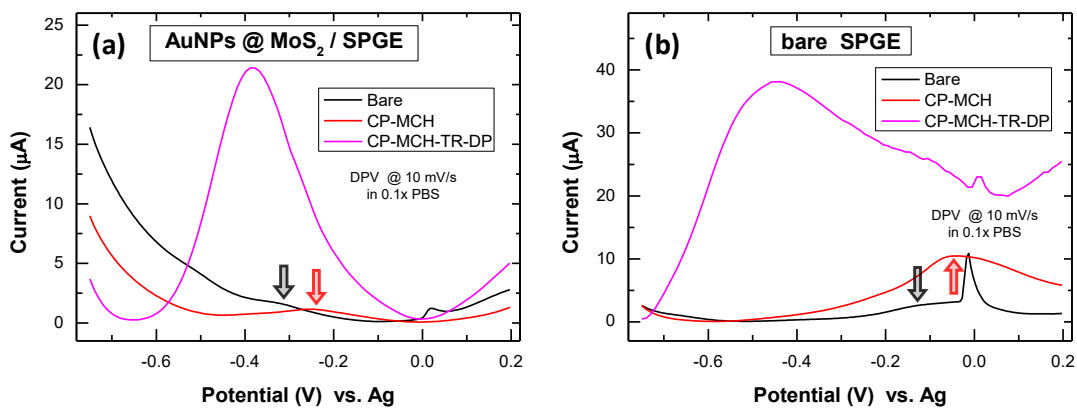

**Figure S9. Monitoring the activity of capture DNA probe (CP) immobilized on bare and modified SPGEs.** (a) DPV profiles confirmed the hybridization of **TR** and **DP** with **CP**, via the redox signal of methylene blue (MB) labeled **DP** on AuNPs@MoS$_2$/SPGE. (b) DPV spectra on the bare SPGE confirmed the immobilization of **CP** and its hybridization with **TR** and **DP**.

**Table S2:** Oligonucleotides used for the confirmation of Immobilization / Hybridization steps

|    | **Oligonucleotides used**         | **Sequence** (from 5′ to 3′)           |
|----|-----------------------------------|----------------------------------------|
| **CP** | Thiol-modified Capture Probe DNA  | 5′-Thiol- TCA ACA TCA GT -3′           |
| **TR** | Target RNA : miRNA-21             | 5′- UAG CUU AUC AGA CUG AUG UUG A -3′  |
| **DP** | **MB**-labeled Detection Probe DNA | 5′- CTG ATA AGC TA -**MB** -3′         |

As an initial step to confirm the hybridization of the capture probe (**CP**) with the target miRNA sequence (**TR**), on the fabricated the AuNPs@MoS$_2$/SPGE platform, "methylene blue (**MB**)" was applied as an electrochemical indicator (denoted as detection probe, **DP**). In other words, in Figure 1, the unlabeled signal amplified ssDNA probe (**AP**) has been replaced by the **MB**-labeled ssDNA probe, (**DP**).

Employing DPV technique, Figure S9a displays the redox characteristics of MB and hence the "hybridization" event at the **CP**-MCH functionalized AuNPs@MoS$_2$/SPGE sensor. For comparison, as shown in Figure S9b, the DPV study was also performed on the bare SPGE sensor, which had also been subjected to the same protocol of surface-modification, immobilization and/or hybridization. It is clear from such a comparison, that the AuNPs@MoS$_2$/SPGE sensor exhibits well-defined and well-resolved voltammograms of MB redox signal, compared to the bare SPGE sensor.

Notably, both bare SPGE and AuNPs@MoS$_2$/SPGE show an obvious peak after the **CP**-MCH functionalization (indicated by red arrows in Figures S9a and S9b). On a careful inspection, it can be found that this peak pre-exists on the unmodified electrodes (indicated by black arrows in Figure S9), prior to **CP**-immobilization, only little shifted to negative potentials. Hence, this DPV signal most probably originates from an intrinsic redox phenomenon related to gold-surface of bare or AuNPs@MoS$_2$ modified SPGEs. Interestingly, **CP**-MCH immobilization led to certain "passivation" effect as evident from the suppression of DPV-background, which in turn results in the prominent appearance of the DPV-signal from Au-surface or AuNPs@MoS$_2$.

## S9. Effect of RuHex Concentration ($C_{RuHex}$).

**Adsorption isotherm of RuHex.**

The influence of $C_{RuHex}$ on CC-detection at **CP**-immobilized electrodes was investigated, as shown in Figure 6. It is observed that $Q_{ad}$ initially increases with $C_{RuHex}$ on bare Au-SPE (SPGE) and AuNPs@MoS$_2$/SPGEs, where the capture DNA probe (**CP**) was immobilized under identical conditions (Figure 6a). The surface excess adsorption isotherm for RuHex, on both bare Au-SPE (SPGE) and AuNPs@MoS$_2$/SPGEs electrodes is plotted as a function of $C_{RuHex}$ in Figure 6b; satisfying the Langmuir adsorption isotherm expression:[8-9]

$$C_{RuHex}/Q_{ad} = 1/KQ_{sat} + (1/Q_{sat}) \times C_{RuHex} \tag{8}$$

where, $Q_{sat}$ is the reduction charge of RuHex with saturated adsorption, $K$ is the association constant of RuHex with DNA.

**Voltammetric responses of RuHex: Effect of $C_{RuHex}$.**

Figure S10 demonstrated electrochemical voltammetric responses as a function of RuHex concentration ($C_{RuHex}$). **CP**-immobilized on MoS$_2$ NSs exhibit a signature response for the redox activity of RuHex, distinctly different from the well-reported **CP**-immobilized Au-sensors.[8-12]

Both the cyclic (CV) and linear-sweep (LSV) voltammetric responses of RuHex, recorded during the adsorption isotherm studies (presented in Figure 6), reveal two kinds of redox activities of RuHex (see the markings in Figure S10). The 1st pair redox peaks (Mark *#1*) originates from RuHex electrostatically bound to the phosphate backbone of DNA (surface-confined redox activity). At sufficiently high $C_{RuHex}$, another set of peaks (Mark **#2**) appears at less-negative potentials, contributing from the RuHex redox probes diffused to the electrode-surface (diffusion-controlled redox activity).[9-12]

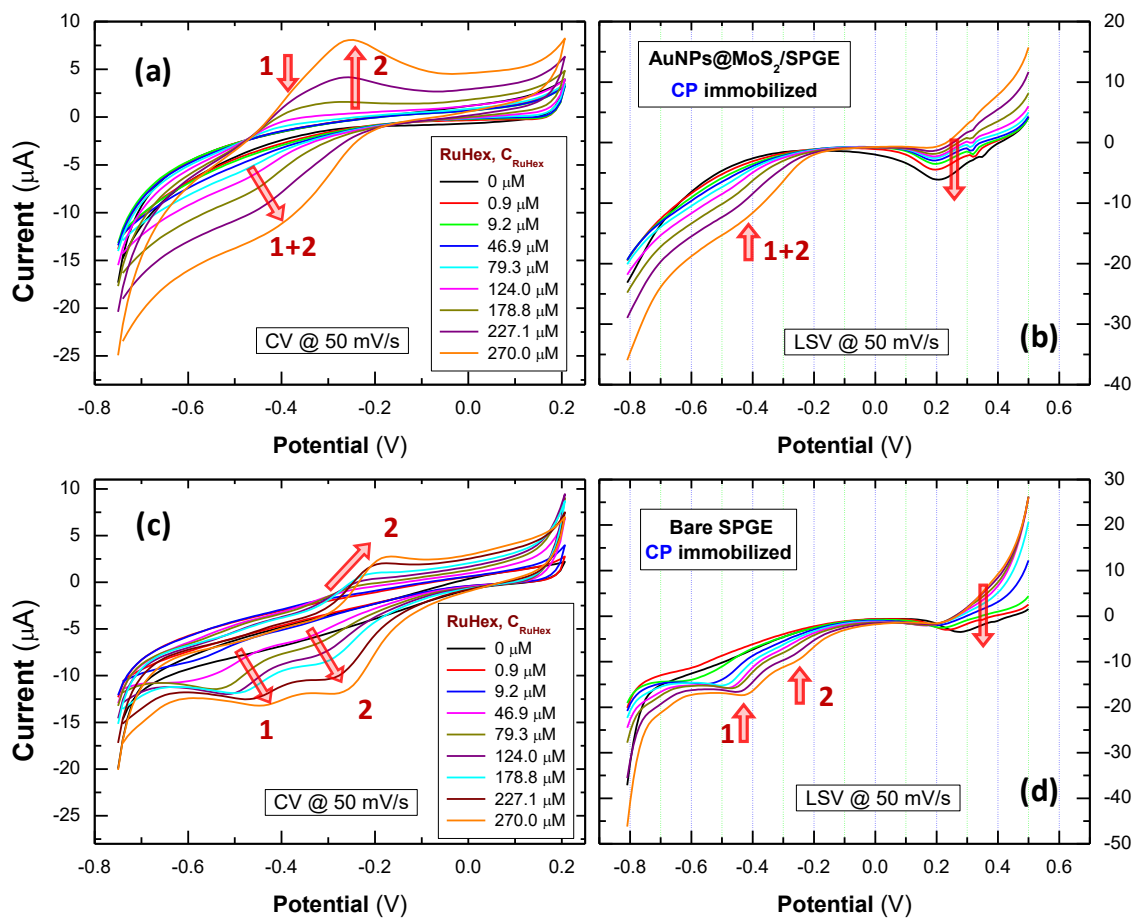

**Figure S10. Electrochemical voltammetric responses as a function of RuHex concentration.** (a) CV and (b) LSV spectra of capture DNA probe (**CP**) immobilized on AuNPs@MoS$_2$/SPGE; and the respective (c) CV and (d) LSV spectra of **CP** immobilized on bare SPGE, recorded in 10 mM TE buffer. $C_{RuHex}$ = RuHex concentration. The MoS$_2$ NSs exhibit signature responses for the redox activity of RuHex, distinctly different from that of well-reported Au-sensors.

At the bare SPGE, these two sets of redox peaks, *#1* and *#2*, are distinctly observed on the cathodic scan. The cathodic peaks (*#1* and *#2*) exhibit blue-shift in potential with increasing $C_{RuHex}$ [Peak *#1*: from -0.6 V (~45 μM) to -0.45 V (~225 μM), and Peak *#2*: from -0.35 V (~45 μM) to -0.3 V (~225 μM)]. The anodic potential-scan demonstrates only the redox peak *#2* that exhibits a blue-shift from -0.3 V (~45 μM) to -0.2 V (~225 μM).

Interestingly, at AuNPs@MoS$_2$/SPGE the two sets of redox peaks, *#1* and *#2*, become noticeable only during the anodic potential-scan, around -0.4 and -0.3 V, respectively. However, peaks *#1* and *#2* overlap during the cathodic potential-scan, exhibiting a peak-potential around -0.46 V. In contrast to bare SPGE, both the redox peaks do not show any noticeable $C_{RuHex}$-dependent shift in peak-potential at the AuNPs@MoS$_2$/SPGE.

Such noticeable dissimilarities in the voltammetric responses of RuHex between the AuNPs@MoS$_2$/SPGE and the bare SPGE could be a sign of different modes of affinity of RuHex at these two electrodes.

Importantly, both CV and LSV responses of RuHex (Figure S10) recommend that, for the CC measurements at the AuNPs@MoS$_2$/SPGEs, the pulse width is needed to extend to 800 mV (applied potential stepping from +0.2 V to -0.6 V), employing 14.5 μM of RuHex to completely neutralize the negative charges of **CP** for efficient quantification. Accordingly, for the CC detection at the bare SPGEs or AuNPs@SPGEs, the pulse width should be kept at 700 mV (from +0.2 V to -0.5 V), using 50 μM of RuHex.

## S10. Optimization of CP–TR–AP Hybridization Strategy.

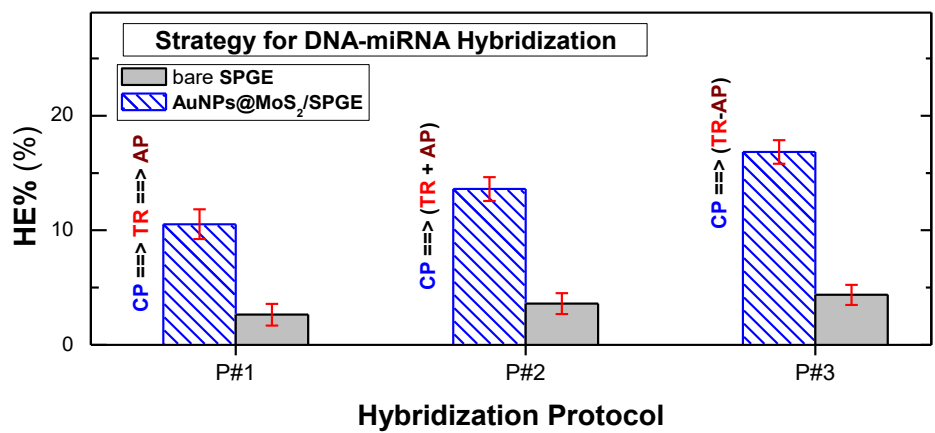

**Figure S11. Optimization of CP–TR–AP Hybridization Strategy.** Hybridization efficiency, $HE\% = (Q_{CP\text{-}TR\text{-}AP} - Q_{CP}) / Q_{CP}$, for different hybridization strategies. Error bars represent the standard deviations estimated from at least three independent measurements.

Table S3: **CP–TR–AP** Hybridization Protocols

| Hybridization Protocol | | 1st Step | 2nd Step | 3rd Step |
|---|---|---|---|---|
| **P#1** | **CP ⇒ TR ⇒ AP** | **CP** Immobilization | **CP–TR** Hybridization | **TR–AP** Hybridization |
| **P#2** | **CP ⇒ (TR + AP)** | **CP** Immobilization | Simultaneous hybridization of **CP** with **TR** and **AP** | -- |
| **P#3** | **CP ⇒ (TR – AP)** | **CP** Immobilization | Hybridization of **CP** with (**TR–AP**) Hybrids | -- |
| | | **TR–AP** Hybridization | | |

## S11. Control Experiment: Confirmation of Amplified function of AP.

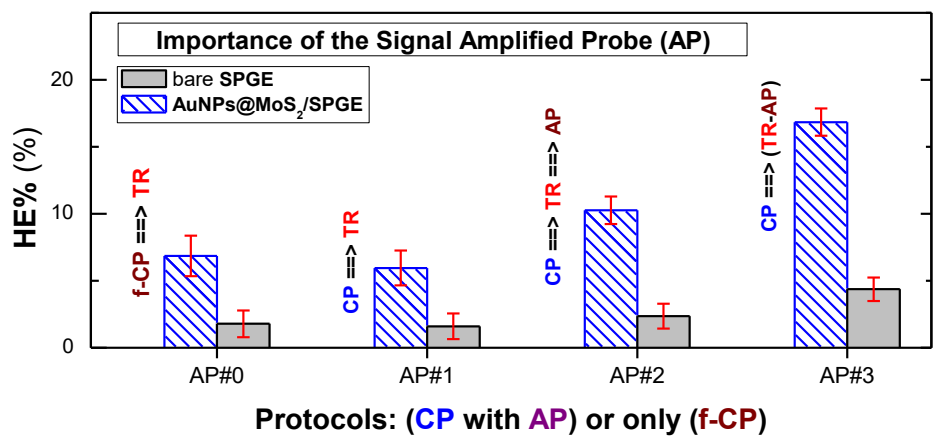

**Figure S12. Importance of the signal amplified probe (AP).** Hybridization efficiency, *HE%*, for different protocols, in absence (only **f-CP**) and presence of **AP** (for **CP**). Error bars represent the standard deviations estimated from at least three independent measurements.

**Table S4:** Oligonucleotides used for control experiment

| | Oligonucleotides used | Sequence (from 5′ to 3′) |
|---|---|---|
| **CP** | Thiol-modified Capture Probe DNA *default choice* | 5′-Thiol- TCA ACA TCA GT -3′ |
| **f-CP** | Thiol-modified Capture Probe DNA *for complete miRNA-hybridization* | 5′-Thiol- TCA ACA TCA GTC TGA TAA GCT A -3′ |
| **TR** | Target RNA : miRNA-21 | 5′- UAG CUU AUC AGA CUG AUG UUG A -3′ |
| **AP** | Signal Amplified Probe DNA | 5′- CTG ATA AGC TA -3′ |

## S12. Selectivity of AuNPs@MoS$_2$/SPGE miRNA Sensor.

**Table S5:** Oligonucleotides used for Selectivity Test

|    |      | **Sequence** (from 5′ to 3′)            |
|----|------|------------------------------------------|
| T1 | TR   | 5′- UAG CUU AUC AGA CUG AUG UUG A -3′   |
| T2 | 1MM  | 5′- U<u>C</u>G CUU AUC AGA CUG AUG UUG A -3′ |
| T3 | 3MM  | 5′- U<u>C</u>G CUU AUC A<u>A</u>A CUG AUG UU<u>C</u> A -3′ |
| T4 | NCT  | 5′- UUA AUG CUA AUC GUG AUA GGG G -3′   |

## S13. Chronocoulometric Responses on CP–TR Hybridization in absence of RuHex.

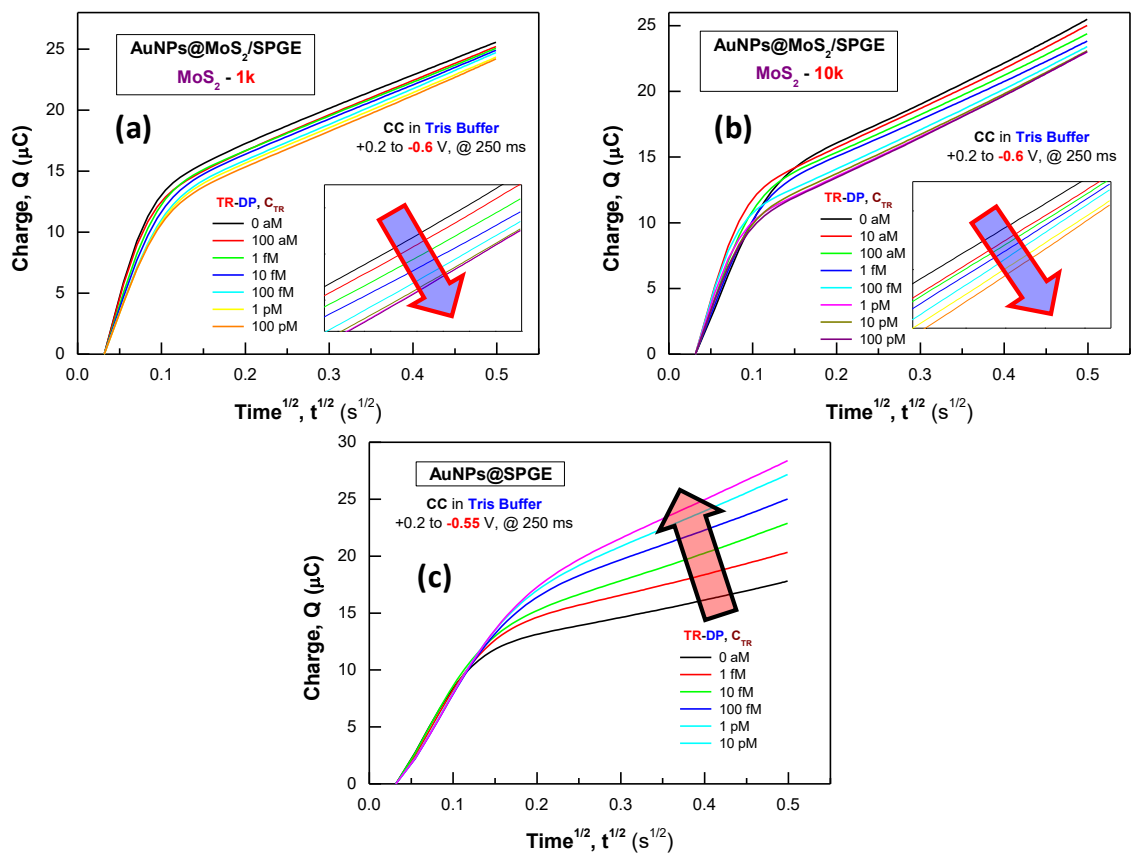

**Figure S13. Effect of $C_{TR}$ on chronocoulometric responses in blank TE buffer (RuHex-free).** Representative CC curves for AuNPs-decorated (a) $MoS_2$(1k), (b) $MoS_2$(10k) and (c) bare SPGE (**AuNPs@SPGE**), recorded in blank 10 mM TE buffer, immediately after the hybridization step at every concentration ($C_{TR}$) of the target miRNA (**TR**), and prior to the addition of RuHex. The $MoS_2$ NSs exhibit an obvious decrease of CC values with $C_{TR}$, distinctly opposite to that exhibited by bare SPGEs.

Interestingly, chronocoulometric responses, recorded in the blank (RuHex-free) TE buffer solution, exhibit an obvious decrease in charge with increasing target miRNA concentration $C_{TR}$, for the AuNPs@MoS$_2$/SPGE sensor (Figure S13). Interestingly, the decreasing trend is more apparent for the sensor comprised of thinner MoS$_2$(10k) NSs (AuNPs@***MoS$_2$(10k)***/SPGE). In contrast, the AuNPs@SPGE sensor exhibits the expected increase of charge as more negatively charged miRNA targets are accumulated on the electrode.

The decrease of the charge at the AuNPs@MoS$_2$/SPGE, upon increasing $C_{TR}$, can be explained through band bending phenomena at semiconductor-electrolyte interface. The AuNPs act as a p-type dopant in MoS$_2$ since the AuCl$_4^-$ ions in solution can strongly withdraw electrons from MoS$_2$ layers and reduce to AuNPs.[13-15] The p-type doping of AuNP in MoS$_2$ was confirmed by XPS. Both Mo$^{4+}$ (Figure 5a) and S$^{2-}$ (Figure 5b) peaks of AuNPs@MoS$_2$ have shifted to lower binding energies compared with that of pure MoS$_2$, indicating a down-shift of the Fermi level in MoS$_2$ due to p-type doping. As a result, the depletion layer width (or band bending) at the metal–semiconductor interface was increased. The depletion layer width (or band bending) is expected to increase further with increasing $C_{TR}$ as more negative charges will develop on the electrode/electrolyte interface; hence the observed charge observed in the CC measurements will progressively decrease.